\documentclass{aa}
\usepackage{txfonts}
\usepackage{graphicx}
\usepackage{natbib}
\usepackage{amssymb}
\usepackage{color}
\bibpunct{(}{)}{;}{a}{}{,}
\begin{document}
\title{On the reliability of mass-loss-rate estimates for AGB stars}
\author{S. Ramstedt\inst{1} \and F. L. Sch\"oier\inst{1,2} \and H. Olofsson\inst{1,2} \and A. A. Lundgren\inst{3}}
\offprints{S. Ramstedt}
\institute{Stockholm Observatory, AlbaNova University Center, SE-106 91 Stockholm, Sweden \\ \email{sofia@astro.su.se} \and Onsala Space Observatory, SE-439 92 Onsala, Sweden \and European Southern Observatory, Casilla 19001, Santiago 19, Chile}    
\date{Received; accepted}
\abstract{In the recent literature there has been some doubt as to the reliability of CO multi-transitional line observations as a mass-loss-rate estimator for AGB stars.}{Using new well calibrated CO radio line observations, the main aim of the work presented here is to carefully evaluate the reliability of CO mass-loss-rate estimates for intermediate- to high-mass-loss-rate AGB stars with different photospheric chemistries.}{Mass-loss rates for 10 intermediate- to high-mass-loss-rate AGB stars are derived using a detailed non-LTE, non-local radiative transfer code based on the Monte-Carlo method to model the CO radio line intensities. The circumstellar envelopes are assumed to be spherically symmetric and formed by constant mass-loss rates. The energy balance is solved self-consistently and the effects of dust on the radiation field and thermal balance are included. An independent estimate of the mass-loss rate is also obtained from the combination of dust radiative transfer modelling with a dynamical model of the gas and dust particles.}{We find that the CO radio line intensities and shapes are successfully reproduced for the majority of our objects assuming a constant mass-loss rate. Moreover, the CO line intensities are only weakly dependent on the adopted micro-turbulent velocity, in contrast to recent claims in the literature. The two methods used in the present work to derive mass-loss-rates are consistent within a factor of $\sim$\,3 for intermediate- to high-mass-loss-rate objects, indicating that this is a lower limit to the uncertainty in present mass-loss-rate estimates. We find a tentative trend with chemistry. Mass-loss rates from the dust/dynamical model are systematically higher than those from the CO model for the carbon stars and vice versa for the M-type stars. This could be ascribed to a discrepancy in the adopted CO/H$_{2}$-abundance ratio, but we caution that the sample is small and systematic errors cannot be excluded.}{}
   
\maketitle
  

\section{Introduction}
\label{s:intro}
Mass loss is the single most important process during the final evolution of low- and intermediate-mass stars on the asymptotic giant branch (AGB) \citep{bloe95}. The mass loss determines the lifetime on the AGB, and thus the maximum luminosity and radius reached \citep[e.g., ][]{marigira07}. Furthermore, since AGB stars are major contributors of mass to the interstellar medium \citep{sedl94,dorshenn95}, and to the integral luminosity of intermediate-age stellar systems, an understanding of the mass-loss phenomenon is also important for studying extra-galactic populations \citep{battdeme04} and for the chemical evolution of galaxies [\cite{schrsedl01}; see also \cite{karalatt07} for a recent paper on yields from AGB stars]. The existence of mass loss is well established \citep{will00}, but its details are less well known. This is unfortunate since even a modest mass-loss-rate change, e.g. by a factor of two, will have a profound effect on the evolution of the star, its nucleosynthesis, and its mass return to the interstellar medium \citep{forechar97}. 

Although mass loss on the AGB can reach rates as high as 10$^{-4}$\,M$_{\sun}$\,yr$^{-1}$, the wind speed is quite low ($\leq$\,30\,km\,s$^{-1}$). The temperature of the outer parts of the atmosphere is low enough for dust to condense. These facts, combined with a high stellar luminosity ($\sim$\,10$^{4}$\,L$_{\odot}$), have lead to the conclusion that the winds of AGB stars are, at least partly, dust-driven, i.e., the mass-loss from the star is driven by radiation pressure on grains and the gas is dragged away from the star by momentum-coupling between the grains and the gas \citep[e.g.,][]{lamecass99}. The effects of the gas-dust interaction on the mass loss and wind formation have been studied by several authors \citep[e.g.,][]{habi96,simietal01}. Time-dependent dynamical models studying the effects of stellar pulsations \citep{bowe88}, and further including detailed time-dependent dust formation \citep{hofnetal95}, have progressed to dynamical models in which the motion of the gas and dust particles is calculated and the radiative transfer is solved \citep{sandhofn03}. These models work well for carbon stars, and support the dust-driven wind scenario, but the formation of silicate grains close enough to the star to drive mass-loss rates of the observed magnitude still poses a problem for M-type stars \citep{woit06,hofnande07}. 

Most likely, the average mass-loss rate increases as the star evolves along the AGB \citep{habi96,schoolof01}. In addition to this steady increase, modulations of the mass-loss rate on shorter time scales are present \citep{habi96}. Indications of this can be seen in observations of circumstellar CO line emission around some AGB stars with double-component line profiles \citep{knapetal98,wintetal03}, and in observations of stellar and ambient Galactic light scattered by dust in circumstellar envelopes (CSEs) \citep{maurhugg00,maurhugg01,maurhugg06}. A process which can have a strong effect on the mass-loss properties of an AGB-star is the He-shell flash (or thermal pulse) \citep{olofetal93,olofetal00,stefscho00,schoetal05b}. Other processes may be
important during the very last stages when the star sheds the
remaining few tenths of a solar mass of its hydrogen envelope (during a few hundred years) before it leaves the AGB \citep{skinetal97}. Finally, the presence of a binary companion may influence the magnitude as well as the characteristics (e.g., the geometry) of the mass loss.

To make progress in the study of AGB stars, it is important to establish reliable mass-loss-rate estimators, which preferably are useable over a wide range of mass-loss rates and time scales, and for different chemical types (M-, S-, and C-stars). Estimating AGB mass-loss rates through observations of circumstellar CO radio line emission, in combination with modelling of the radiative transfer, is considered to be one of the most reliable methods. CO line emission has also been used in a number of studies aimed at tracing the mass-loss history \citep{schoetal02,kempetal03,teysetal06,decietal07}, 
and at identifying drastic mass-loss-rate changes at the end of the AGB \citep{hesketal90,justetal96}.
Detailed studies of the mass-loss properties of carbon stars \citep{schoolof01,schoetal02}, M-type AGB stars \citep{olofetal02,delgetal03}, and S-type AGB stars \citep{ramsetal06} have been performed on samples dominated by low-mass-loss-rate
objects. It was concluded that for this type of objects the CO radio
line emission is a good mass-loss-rate estimator, although not
entirely free of problems. 

It has been known for some time that for the higher mass-loss-rate objects the situation is less clear \citep{saha90,kast92,schoolof01}. A higher mass-loss rate leads to a decrease in the drift velocity between the dust and the gas and therefore a less efficient heating of the gas through collisions with dust grains.
Since CO line emission is also a dominant coolant, the effect of an
increased mass-loss rate is increased cooling of the circumstellar gas. This, in combination with saturation of the spectral lines, leads to a
situation where the integrated CO line intensities are less dependent on
the mass-loss rate when it exceeds about
10$^{-5}$\,M$_{\odot}$\,yr$^{-1}$; the higher the frequency of the
lines the lower the mass-loss rate when the effect sets in. This
makes CO line emission a questionable mass-loss-rate estimator at mass-loss
rates in excess of 10$^{-5}$\,M$_{\odot}$\,yr$^{-1}$, an unfortunate
effect since we know that mass-loss rates up to
10$^{-4}$\,M$_{\odot}$\,yr$^{-1}$ are reached during the very final
stage, and this is also when the stars make their largest
contribution to the return of nuclear-processed material to the interstellar medium. In addition, \citet{kempetal03} concluded that standard circumstellar radiative transfer models could not explain some CO line intensity ratios, and that the mass-loss-rate estimate is sensitively dependent on the (unknown) local turbulent line width.

An alternative way to estimate the mass-loss rate, not subject to the same problems, is to combine modelling of the dust radiative transfer with solving the coupled equations of motion for the gas and the dust particles \citep{habietal94}. The gas expansion velocity is a strong function of the dust-to-gas mass-loss-rate ratio and therefore a solution of the kinematical equations that fits the gas expansion velocity and the dust optical depth, obtained through dust radiative transfer modelling, results in an independent estimate of the mass-loss rate.

Presented here are a detailed radiative transfer modelling of primarily new, well-calibrated, and high signal-to-noise ratio observations of CO  $J$\,=\,1\,$\rightarrow$\,0, 2\,$\rightarrow$\,1, 3\,$\rightarrow$\,2, 4\,$\rightarrow$\,3, and 6\,$\rightarrow$\,5 emission from a sample of 10 well-studied high-mass-loss-rate AGB stars, and the resulting mass-loss-rate estimates. In addition, dust emission modelling combined with a dynamical model give independent estimates of the mass-loss rates. The dust emission modelling also provides the dust information necessary for the CO modelling, e.g., the dust radiation field. By obtaining the mass-loss rates for the stars using two independent methods, we can evaluate the reliability of the mass-loss-rate estimators, and also look for trends with e.g. mass-loss rate and chemistry. A similar investigation comparing mass-loss-rates derived from ISO-SWS observations of the silicate dust features with mass-loss rates from CO modelling found in the literature on a sample of low-mass-loss-rate M-type AGB stars has been performed by \citet{herahony05}.

\begin{table}
\caption{The sample with spectral types, variable types, periods ($P$), luminosities ($L_{\star}$), and distances ($D$).}
\label{sample}
$$
\begin{array}{p{0.3\linewidth}ccccccc}
\hline
\noalign{\smallskip}
\multicolumn{1}{l}{{\mathrm{Source}}} &
\multicolumn{1}{c}{{\mathrm{Spec.}}} &
\multicolumn{1}{c}{{\mathrm{Var.}}} &&
\multicolumn{1}{c}{{\mathrm{P}}} &
\multicolumn{1}{c}{L_{\star}} && 
 \multicolumn{1}{c}{D}\\ 
&
\multicolumn{1}{c}{\mathrm{Type}} &
\multicolumn{1}{c}{\mathrm{Type}} &&
\multicolumn{1}{c}{[\mathrm{days}]} &
\multicolumn{1}{c}{[\mathrm{L}_\odot]} &&
\multicolumn{1}{c}{[\mathrm{pc}]} 
\\
\noalign{\smallskip}
\hline
\noalign{\smallskip}
\object{LP And} & \mathrm{C} & \mathrm{Mira} &&  620 & \phantom{0}8200 && \phantom{1}620 \\
\object{CW Leo} & \mathrm{C} & \mathrm{Mira} && 630 & \phantom{0}8300 && \phantom{1}120 \\
\object{RW LMi} & \mathrm{C} & \mathrm{SRa} && 640 & \phantom{0}8500 && \phantom{1}390 \\
\object{V384 Per} &  \mathrm{C} & \mathrm{Mira} && 535 & \phantom{0}7000 && \phantom{1}590 \\
\object{AFGL 3068} & \mathrm{C} & \mathrm{Mira} && 696 & \phantom{0}9300 && 1090 \\	
 & & & && && \\
\object{TX Cam} & \mathrm{M} & \mathrm{Mira} && 557 & 11900 && \phantom{1}440 \\
\object{GX Mon} & \mathrm{M} & \mathrm{Mira} && 527 & 11100 && \phantom{1}630 \\
\object{WX Psc} & \mathrm{M} & \mathrm{Mira} && 660 & 14600 && \phantom{1}720 \\	
\object{IK Tau}	& \mathrm{M} & \mathrm{Mira} && 500 & 10400 && \phantom{1}300 \\
\object{IRC--10529}	& \mathrm{M} & \mathrm{Mira} && 680 & 15100 && \phantom{1}890 \\
\noalign{\smallskip}
\hline
\end{array}
$$
\end{table}

The sample and the observational data are presented in Sects~\ref{s:sample} and \ref{s:obsdata}. In Sect.~\ref{s:circum} the adopted circumstellar model is discussed. In Sect.~\ref{s:COmod} the CO radiative transfer model is described and CO as a mass-loss rate estimator is discussed. In Sect.~\ref{s:dynmod}, we describe how dust radiative transfer and the dynamical model is used to estimate mass-loss rates. The results are presented in Sect.~\ref{s:res}, and outlined for each individual object in Sect.~\ref{s:ind}. In Sect.~\ref{s:discuss} we discuss and compare the two methods, and in Sect.~\ref{s:conc} we give our conclusions.


\section{The sample}
\label{s:sample}
The sample, presented in Table~\ref{sample}, consists of 10 well-studied AGB stars, with intermediate to high mass-loss rates. They are chosen with a spread in mass-loss rate, to allow studies of possible trends. Dependences on chemistry can also be investigated, since half of the stars were chosen to have a photospheric C/O-ratio $>$1 (carbon stars) and the other half C/O$<$1 (M-type).

Bolometric luminosities are estimated using the period-luminosity relation of \cite{feasetal06} for the carbon stars, and that of \cite{whitetal94} for the M-type stars. Periods are mainly taken from the Combined General Catalogue of Variable stars \citep{samuetal04}. The periods for LP And, AFGL 3068, and IK Tau are from \citet{mensetal06}, \citet{lebertre92}, and \citet{delgetal03}, respectively. Distances are derived by fitting the spectral energy distribution (SED) calculated with DUSTY (Sect.~\ref{s:dustmod}) to observed fluxes from the near-IR to the mm-range (Sect.~\ref{s:flux} and Appendix~\ref{a:flux}).

\begin{table}
\caption{Main-beam efficiencies ($\eta_{\mathrm{mb}}$), and main-beam FWHM:s ($\theta_{\mathrm{mb}}$), at the frequencies of the CO lines.}
\label{efficiencies}
$$
\begin{array}{ccccccccc}
\hline
\noalign{\smallskip}
\multicolumn{1}{c}{{\mathrm{Transition}}} &&
\multicolumn{1}{c}{{\mathrm{Frequency}}} &&
\multicolumn{1}{c}{\eta_{\mathrm{mb}}}  && 
\multicolumn{1}{c}{\theta_{\mathrm{mb}}} &
\multicolumn{1}{c}{{\mathrm{Telescope}}} \\ 
&&
\multicolumn{1}{c}{{\mathrm{[GHz]}}} && && 
\multicolumn{1}{c}{{\mathrm{['']}}} & \\
\noalign{\smallskip}
\hline
\noalign{\smallskip}
J=1\rightarrow0 && 115.271     && 0.43 && 33 & \mathrm{OSO} \\
J=2\rightarrow1 && 230.538    && 0.66 && 21 & \mathrm{JCMT} \\
J=3\rightarrow2 && 345.796    && 0.62 && 14 & \mathrm{JCMT} \\	
J=4\rightarrow3 && 461.041    && 0.5\phantom{0} && 11 & \mathrm{JCMT} \\	
J=6\rightarrow5 && 691.473    && 0.3\phantom{0}  && 8 & \mathrm{JCMT}\\	
\noalign{\smallskip}
\hline
\end{array}
$$
\end{table}
\begin{table}[t]
\caption{Integrated CO($J\rightarrow J-1$) intensities, $\int T_{\mathrm{mb}}\,dv$, in K\,km\,s$^{-1}$.}
\label{intensities1}
$$
\begin{array}{p{0.27\linewidth}cccccccc}
\hline
\noalign{\smallskip}
\multicolumn{1}{l}{{\mathrm{Source}}} &
\multicolumn{1}{c}{{\mathrm{OSO}}} &&
\multicolumn{4}{c}{{\mathrm{JCMT}}} \\ 
\cline{2-2}
\cline{4-7}
&
\multicolumn{1}{c}{1\rightarrow0} &&
\multicolumn{1}{c}{2\rightarrow1} &
\multicolumn{1}{c}{3\rightarrow2} &
\multicolumn{1}{c}{4\rightarrow3}& 
\multicolumn{1}{c}{6\rightarrow5} \\
\noalign{\smallskip}
\hline
\noalign{\smallskip}
\object{LP And} & \phantom{0}63 && 104 &  \phantom{0}124\phantom{^a} &  \phantom{0}143\phantom{^a} & 139 \\
\object{CW Leo}	 & 386 && 689  & 1070^a  & 1230^a & \cdots \\
\object{RW LMi} &108 && 123 &  \phantom{0}244^a &  \phantom{0}246^a & \cdots \\
\object{V384 Per} & \phantom{0}35 &&  \phantom{0}63 & \phantom{00}40\phantom{^a}  & \phantom{00}79\phantom{^a} & \cdots  \\
\object{AFGL 3068} & \phantom{0}48 &&  \phantom{0}72  &  \phantom{00}73\phantom{^a} &  \phantom{00}93\phantom{^a} &  \phantom{0}55 \\	
 & && & & & \\
\object{TX Cam} & \phantom{0}20&& \phantom{0}61  & \phantom{00}71\phantom{^a}    &\phantom{0}149\phantom{^a} & \cdots \\
\object{GX Mon}& \phantom{0}31&&  \phantom{0}61 &  \phantom{00}80\phantom{^a}  & \phantom{00}79\phantom{^a} & \cdots \\
\object{WX Psc} & \phantom{0}44&& \phantom{0}81 &  \phantom{00}70\phantom{^a} & \phantom{00}84\phantom{^a}  & \phantom{0}23 \\	
\object{IK Tau}	& \phantom{0}46 && \phantom{0}95  & \phantom{0}125^a &\phantom{0}130\phantom{^a} & \cdots \\
\object{IRC--10529}	& \phantom{0}16 && \phantom{0}45 & \phantom{0}27  & \phantom{00}22\phantom{^a} & \cdots \\
\noalign{\smallskip}
\hline
\end{array}
$$
$^a$ JCMT archival data
\end{table}
%

\section{Observational data}
\label{s:obsdata}
\subsection{New observations of CO radio line emission}
\label{s:new}
When estimating the physical properties of the CSEs, e.g., mass-loss rates and kinetic temperatures, it is very important to have well-calibrated data for all lines \citep{schoolof01}. CO $J$\,=\,1\,$\rightarrow$\,0 (115.271 GHz) data was obtained in December 2003 using the Onsala Space Observatory (OSO) 20\,m telescope\footnote{The Onsala 20\,m telescope is operated by the  Swedish National Facility for Radio Astronomy, Onsala Space observatory at Chalmers University of technology}. The James Clerk Maxwell Telescope\footnote{The JCMT is operated by the Joint Astronomy Centre in Hilo, Hawaii on behalf of the
present organizations: the Particle Physics and Astronomy Research Council in the 
United Kingdom, the National Research Council of Canada and the Netherlands
Organization for Scientific Research.}  (JCMT) located on Mauna Kea was used to collect CO $J$\,=\,2\,$\rightarrow$\,1 (230.538 GHz), $J$\,=\,3\,$\rightarrow$\,2 (345.796 GHz), $J$\,=\,4\,$\rightarrow$\,3 (461.041 GHz), and $J$\,=\,6\,$\rightarrow$\,5 (691.473 GHz) line emission from the sample sources. The observations were performed in June, July, and October 2003.  

All observations were made in a dual-beamswitch mode, 
where the source is alternately placed in the signal and the reference
beam, using a beam throw of about $2\arcmin$ at JCMT and $11\arcmin$ at OSO. 
This method produces very flat baselines. Regular pointing checks made on strong continuum sources (JCMT) 
or SiO masers (OSO) were found to be consistent with the pointing model within $\approx$\,3$\arcsec$. 

The data was reduced in a standard way using XS\footnote{XS is a package developed by P. Bergman to reduce and analyze a large number of single-dish spectra. It is publicly available from {\tt ftp://yggdrasil.oso.chalmers.se}} by removing a first order polynomial fitted to the emission-free channel ranges, and then binned in order to improve the signal-to-noise ratio. The intensity scale is given in main-beam-brightness temperature, $T_{\mathrm{mb}}$\,=\,$T_{\mathrm{A}}^{\star}/\eta_{\mathrm{mb}}$, where $T_{\mathrm A}^{\star}$ is the
antenna temperature corrected for atmospheric attenuation using the
chopper-wheel method, and $\eta_{\mathrm{mb}}$ the main-beam
efficiency. The adopted main-beam efficiencies are given in Table~\ref{efficiencies}.
The uncertainty in the absolute intensity scale is estimated to be about $\pm 20$\% for all lines except the CO $J$\,=\,6\,$\rightarrow$\,5 lines, where we estimated a higher uncertainty of about $\pm 30$\%. 

The observed spectra are presented in Figs~\ref{lpand_mod}, \ref{txcam_mod}, and \ref{cwleo_mod}\,--\,\ref{irc10529_mod}, and velocity-integrated intensities are reported in Table~\ref{intensities1}.

\subsection{Dust continuum emission}
\label{s:flux}
The SEDs are constructed from JHKLM-band, sub-millimetre, and millimetre photometric flux densities from the literature, together with IRAS data. SCUBA data (at 450 and 850 $\mu$m) was downloaded from the JCMT archive for CW Leo, AFGL 3068, and RW LMi. All fluxes and references are presented in Appendix~\ref{a:flux}.

To check the dust radiative transfer model results, IRAS and ISO spectroscopic data were used. IRAS-LRS spectra were downloaded\footnote{http://www.iras.ucalgary.ca/$\sim$volk/getlrs\_plot.html} for all stars in the sample. An absolute calibration correction was applied according to \citet{volkcohe89} and \citet{coheetal92}. The ISO LWS and SWS Highly Processed Data Products (HPDP) were downloaded and not processed further. LWS spectra were available for all stars, except GX Mon. SWS spectra were found for TX Cam and WX Psc (SWS spectra for carbon stars were not searched for). For IK Tau, only a grating scan covering the wavelength range $~$29--49 $\mu$m was available, and for IRC--10529, the SWS spectrum is very noisy above 28 $\mu$m.


%
\section{The circumstellar model}
\label{s:circum}
The CSE is assumed to be spherically symmetric and formed by a constant mass-loss rate. It is assumed to be expanding at a constant velocity, derived from fitting the CO line widths, and to have a micro-turbulent velocity distribution with a doppler width of 0.5\,km\,$\mathrm{s^{-1}}$ (e-folding value) throughout. In addition, a thermal contribution to the local line width is added, based on the derived kinetic temperature of the gas. The density structure is obtained from the conservation of mass.

Even in this simple model there are a number of uncertainties that will affect the strengths and the shapes of the CO lines and the shape of the SED, e.g., the thermodynamics of the gas, the CO chemistry, the dust composition, the optical properties of the dust, and the grain size distribution. In addition, there can be major deviations from this model due to e.g. non-isotropic, time-variable mass loss, and a clumpy medium. Finally, the objects are long-period variables and the data are sometimes taken at different epochs. In many cases it may be impossible to disentangle which of these assumptions and effects, if any, are affecting the emerging radiation. This taken into account, the most reasonable approach is to choose the simplified circumstellar model and analyze all stars using the same method.

\section{CO line modelling}
\label{s:COmod}
Only a short description of the radiative transfer analysis of the circumstellar CO radio line emission is given here, since the adopted model has been described in detail in previous articles [\cite{schoolof01} for carbon stars, and \cite{olofetal02} for M-type stars].

The non-LTE radiative transfer code is based on the Monte Carlo method. 41 rotational levels are included in each of the two lowest vibrational states (v=0 and v=1) in the excitation analysis of the CO molecule. Energy levels and radiative transition probabilities, as well as collisional rate coefficients for collisions between CO and $\rm{H_{2}}$, are taken from \cite{schoetal05}\footnote{http://www.strw.leidenuniv.nl/$\sim$moldata}. An ortho-to-para ratio of 3 is adopted when weighting together collisional rate coefficients for CO in collisions with ortho- and para-H$_{2}$.
Radiation from the central source (assumed to be a blackbody), thermal dust grains (Sect.~\ref{s:dustmod}), and the cosmic microwave background is included. 

The energy balance equation for the gas is solved self-consistently, by including line cooling from CO and $\mathrm{H_{2}}$ and cooling due to the adiabatic expansion of the gas. The dominant gas-heating mechanism is collisions between the $\mathrm{H_{2}}$ molecules and the dust grains. Photoelectric heating is also included and it is mostly important in the outer parts of the CSE. The free parameters describing the dust are combined in the parameter, $h$, defined as

\begin{equation}
\label{h}
h = \left( \frac{\Psi}{0.01} \right) \left( \frac{2.0 \  \mathrm{g \ cm^{-3}}}{\rho_{\mathrm{g}}} \right) \left( \frac{0.05 \ \mu \mathrm{m}}{a_{\mathrm{g}}} \right),
\end{equation}

\noindent
where $\Psi$ is the dust-to-gas mass-loss-rate ratio and $\mathrm{\rho_{g}}$ and $a_{\mathrm{g}}$ are the average density and radius of an individual dust grain, respectively. Following \citet{schoolof01}, we have adopted an average efficiency factor for momentum transfer (see Sect.~\ref{ss:dynmod}), which is constant throughout the CSE, $\langle Q_{\rm{rp}} \rangle^{\dagger}$=3\,$\times$\,10$^{-2}$. However, the particular choice of $\langle Q_{\rm{rp}} \rangle^{\dagger}$ is not so critical. As long as the product $h \langle Q_{\rm{rp}} \rangle^{\dagger 3/2}$ is kept constant (and h is a free parameter), the model results are essentially the same for intermediate and high mass-loss-rate objects, for a given $\dot{M}$ (see Eqns (4) and (6) in \citet{schoolof01}).

The radial distribution of CO is estimated using the model presented in Mamon et al. (1988). It includes photodissociation, shielding, and chemical exchange reactions. The photospheric abundance of CO relative to $\mathrm{H_{2}}$, is assumed to be 1\,$\times$\,10$^{-3}$ for the carbon stars \citep{zuckdyck86} and 2\,$\times$\,10$^{-4}$ for the M-type stars \citep{kahajura94}. The luminosity and the inner radius of the CSE were taken from the SED fitting. It was shown in \citet{schoolof01} that, in the intermediate- to high-mass-loss-rate stars, the CO radio line intensities are essentially unaffected by a change in these parameters. The main reason is that in the higher-mass-loss-rate CSEs, CO is collisionally excited out to large radii. The mass-loss rate and $h$ are the remaining free parameters in the CO line modelling.


\section{Dust radiative transfer and dynamical model}
\label{s:dynmod}

\subsection{Dust emission modelling}
\label{s:dustmod}
Dust radiative transfer through a spherically symmetric envelope possesses scaling properties \citep{ivezelit97}.
This fact is put to use in the publicly available radiative transfer code 
DUSTY\footnote{http://www.pa.uky.edu/$\sim$moshe/dusty/} 
adopted here to model the observed continuum emission. 
The scaling properties of the solution makes DUSTY ideal for studying a large sample of stars, since
only one model grid, although large, needs to be calculated for each dust type. 

The most important parameter in the dust radiative transfer modelling is the dust optical depth
\begin{equation}
\label{tau}
\tau_{\lambda} = \kappa_{\lambda} \int^{r_{\mathrm e}}_{r_{\mathrm i}} \rho_{\mathrm d}(r) dr,
\end{equation} 
where $\kappa_{\lambda}$ is the dust opacity per unit mass, $\rho_{\mathrm d}$ the dust mass density, and 
the integration is made from the inner ($r_{\mathrm i}$) 
to the outer ($r_{\mathrm e}$) radius of  the CSE. Prompt dust formation is assumed 
at the inner radius. Amorphous carbon dust grains, with the optical constants presented in 
\citet{suh00}, are adopted for the carbon stars, and amorphous silicate grains are adopted for the M-type stars \citep{justtiel92}. 
For simplicity, the dust grains are assumed to be of the same size with a radius of 0.1\,$\mu$m, and have a density of 2\,g\,$\mathrm{cm^{-3}}$ (carbon grains) or 3\,g\,cm$^{-3}$ (silicate grains).

In the modelling, the dust optical depth, specified at 10\,$\mu$m, is varied in the range 0.1$-$10 in steps of $\sim$10\%. In addition, the dust temperature at the inner radius of the dust envelope $T_{\mathrm d}(r_{\mathrm i})$, and the stellar effective temperature $T_{\mathrm{eff}}$, are varied.
These are the three adjustable parameters in the analysis. 
The SED model result depends only weakly on the other input parameters, which are fixed at reasonable values. 
The size of the envelope is fixed at $r_{\mathrm e}$/$r_{\mathrm i}$=20000 to make sure that
the whole envelope extent determined from the CO emission [$r_\mathrm{ p}$; see Table~3 in 
\citet{schoolof01}] is covered. Using an even larger size does not affect the result.

The temperature at the inner radius should be close to (or lower than) the dust condensation temperature. In the modelling it is varied in the range 300$-$1500\,K, in steps of 100\,K.
The canonical dust condensation temperature of carbonaceous dust grains is $\approx$1500\,K, while the condensation temperature for the silicate grains is expected to be somewhat lower.
However, prompt dust formation is an oversimplification. Given the complex chemistry close to the star, where shocks are present and grain growth is important, the dust-to-gas ratio probably depends on the radius. The dust temperature at the inner radius should therefore be treated as a measure of the characteristic radius where dust formation is essentially completed.
The stellar effective temperature is varied in the range 1800$-$2400\,K (also in steps of 100\,K), 
characteristic of high-mass-loss-rate AGB stars, but lower than that of AGB stars with optically thin dust envelopes (e.g., Lambert et~al.\ 1986\nocite{lambetal86};
Bergeat et~al.\ 2001\nocite{bergetal01}; Heras\,\&\,Hony\,2005\nocite{herahony05}).

When scaling the result from DUSTY, the interstellar extinction is taken into account. It is calculated from the galactic longitude for each star, using equation (8) in \cite{groeetal92}. 

\subsection{Dynamical model}
\label{ss:dynmod}
When light is scattered by a dust particle, some of its momentum is given to the scattering particle. For particles with an arbitrary shape, the efficiency factor for this momentum transfer, depends on the orientation of the particle and on the polarization of the light. However, for spherical particles, as assumed here, it is independent of both. The dimensionless efficiency factor is given by the momentum fraction taken from the incident beam, divided by the geometrical cross section of a particle:

\begin{equation}
\label{Q}
Q_{\lambda}=\frac{4}{3} \kappa_{\lambda} a_{\mathrm{g}} \rho_{\mathrm{g}}
\end{equation}

In the CSEs of AGB stars, momentum transfer from photons to dust grains pushes the dust outward, and friction between the dust and gas particles determines the gas outflow velocity. 
By assuming that the wind is entirely dust-driven, conservation of mass and momentum 
can be used to set up equations for the velocity of the dust ($v_{\rm{d}}$) and the gas ($v_{\rm{g}}$)
particles,  

\begin{equation}
\label{v_dust}
\frac{dv_{\mathrm{d}}^{2}}{dr} = \frac{2}{r^2} \bigg[ \frac{\sigma_{\mathrm{d}} \langle Q_{\mathrm{rp}} \rangle L_{*}}{4 \pi c m_{\mathrm{d}}} - GM_{*} - \frac{\sigma_{\mathrm{d}} \dot{M} v_{\mathrm{dr}} \sqrt{(v_{\mathrm{s}}^{2} + v_{\mathrm{rel}}^{2})}}{4 \pi m_{\mathrm{d}} v}\bigg]
\end{equation}

\begin{equation}
\label{v_gas}
\frac{dv_{\rm{g}}^{2}}{dr} = \frac{2}{r^2} \bigg[ -GM_{*} + \frac{\Psi \sigma_{\mathrm{d}} \dot{M} v_{\mathrm{dr}} \sqrt{(v_{\mathrm{s}}^{2} + v_{\mathrm{rel}}^{2})}}{4 \pi m_{\mathrm{d}} v_{\mathrm{d}}}\bigg],
\end{equation}

\noindent
where $\sigma_{\mathrm{d}}$ and $m_{\mathrm{d}}$ are the collisional cross section and mass of a dust grain, $\langle Q_{\mathrm{rp}} \rangle$ the radiation pressure efficiency averaged over the photon spectrum, $L_{*}$ and $M_{*}$ the stellar luminosity and mass, $v_{\rm{dr}}$ (=$v_{\rm{d}}-v_{\rm{g}}$) the drift velocity between the dust and the gas, and $v_{\mathrm{s}}$ the sound speed assumed to be 2\,km\,$\rm{s}^{-1}$ \citep{lamecass99}. A stellar mass of 1.5\,M$_{\odot}$ is used for all stars in the sample \citep{kahaetal00,olivetal01}.

The radiation pressure efficiency $Q_{\lambda}$ is determined and averaged over the photon spectrum, 
 
\begin{equation}
\label{Qmean}
\langle Q_{\rm{rp}} \rangle (r) = \frac{\int Q_{\lambda} B_{\lambda} e^{-\tau_{\lambda}} \mathrm{d}\lambda}{\int B_{\lambda} e^{-\tau_{\lambda}} \mathrm{d}\lambda },
\end{equation}

\noindent
where $B_{\lambda}$, calculated from the effective temperature of the star, is weighted by the absorption determined by the optical depth $\tau_{\lambda}$ derived using the dust opacity and the dust density structure from the dust emission modelling (Eqn.~\ref{tau}). This includes absorption, but not re-emission at longer wavelengths. This simplification will probably not have a big effect on the result due to the wavelength dependence on the radiation pressure efficiency. DUSTY includes an option where the wind structure can be calculated by solving the hydrodynamical equations. The drawback compared to the dynamical model presented here is that a dust-to-gas ratio needs to be assumed in order to get an estimate of the mass-loss rate, but as opposed to our model, it includes the re-emission by the dust. A comparison between the results from our dynamical model and the mass-loss-rate estimates derived by DUSTY on five of our sources showed that both methods give the same result (within $\sim$\,30\%) for the same dust-to-gas ratio.
 
$\langle Q_{\rm{rp}} \rangle$ rapidly decreases and reaches its terminal value, $\langle Q_{\rm{rp}} \rangle^{\dagger}$, at $\approx\,20-30\,R_{\star}$. Equations (\ref{v_dust}) to (\ref{Qmean}) are solved in an iterative way, for different mass-loss rates, until the dust density structure is consistent with the optical depth derived in the dust radiative transfer model, and a fit to the observed gas expansion velocity is found. In this way, an independent estimate of the mass-loss rate is obtained. 

\begin{figure}[h]
\centering{   
\includegraphics[width=9cm,height=7cm]{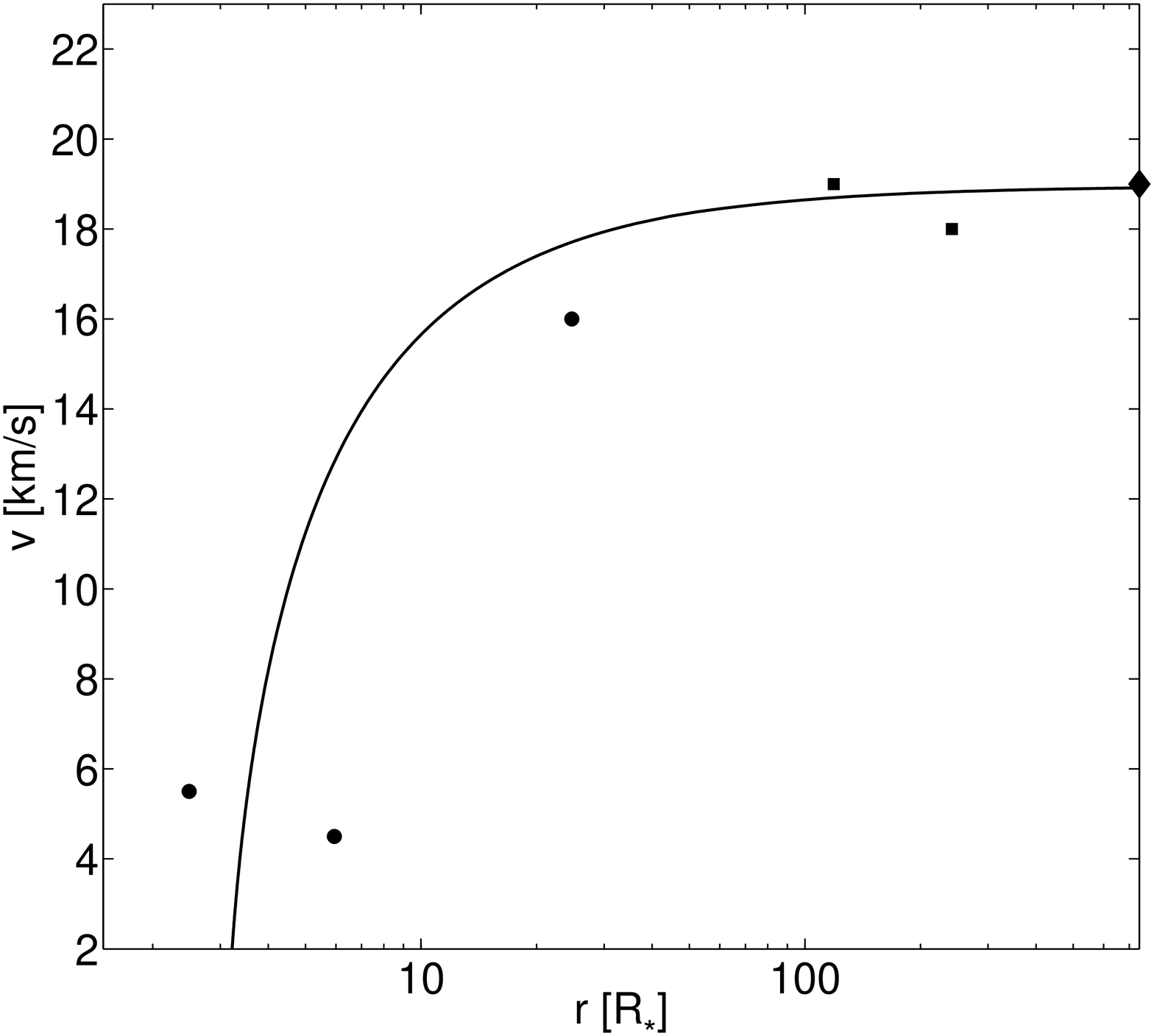}
\caption{The velocity profile for IK Tau from the dynamical model overlayed with the velocity data obtained from mapping of maser emission \citep[see][and references therein]{bainetal03}. H$_{2}$O masers (circles) and OH maser (squares) have been used. The expansion velocity derived from the CO data is marked by a diamond.}
\label{bain}}
\end{figure}

\citet{bainetal03} investigated maser emission towards four AGB stars and derived velocity profiles for their circumstellar envelopes. In Fig.~\ref{bain}, our result for IK Tau is overlayed and shown to be consistent with their data.

\section{Results}
\label{s:res}
\subsection{CO line modelling}
\label{s:res:COmod}
\subsubsection{Strategy for finding the best-fit model}
\label{s:COstrat}
Only two parameters are adjustable when fitting the model to the observed integrated line intensities: the mass-loss rate, ${\dot{M}}$, and the $h$-parameter. Models are calculated for a large number of mass-loss rates and $h$-parameters for each star and the best-fit model is found using $\chi^{2}$-statistics. The reduced $\chi^{2}$ for the best-fit model is estimated from

\begin{equation}
\label{chi2_red}
\chi_{\mathrm{red}}^{2} = \frac{\chi_{\mathrm{min}}^{2}}{N-p},
\end{equation}

\noindent
where $N$ is the number of observational constraints, and $p$ the number of adjustable parameters (2 in our case). $\chi_{\mathrm{min}}^{2}$ is the minimum $\chi^{2}$ obtained in the model runs, which is calculated as,

\begin{equation}
\label{chi2}
\chi^{2} = \sum_{i=1}^{N} \frac{(I_{\mathrm{mod,i}}-I_{\mathrm{obs,i}})^{2}}{\sigma_{i}^{2}},
\end{equation}

\noindent
where $\sigma$, for the high-quality data presented here, is dominated by the calibration uncertainties assumed to be 20\% for all lines except the $J$=6$\rightarrow$5 lines where they are assumed to be 30\% (see Sect.~\ref{s:new}).
\begin{table}
\caption{Mass-loss rates, $h$-parameters, and terminal gas expansion velocities $v_{\rm{e}}$, derived by fitting the CO line radiative transfer model to the observations.}
\label{mdotCO}
$$
\begin{array}{p{0.3\linewidth}cccccccc}
\hline
\noalign{\smallskip}
\multicolumn{1}{l}{{\mathrm{Source}}} &
\multicolumn{1}{c}{{\dot{M}}} && 
\multicolumn{1}{c}{{h}} &&
\multicolumn{1}{c}{{v_{\mathrm{e}}}} &&
\multicolumn{1}{c}{{\mathrm{\chi^{2}_{red}}}} &
\multicolumn{1}{c}{{N}} \\ 
 &
\multicolumn{1}{c}{[\mathrm{10^{-6}\ M_{\odot}\ yr^{-1}}]} &
 &&& \multicolumn{1}{c}{[\mathrm{km\ s^{-1}}]} & & \\
\noalign{\smallskip}
\hline
\noalign{\smallskip}
\object{LP And} & \phantom{0}7 && 2.0 && 14.0 && 0.5 & 5 \\
\object{CW Leo} & 20 && 1.0 && 14.5 && 0.1 & 4 \\
\object{RW LMi} & \phantom{0}5 && 1.5 && 17.0 && 0.7 & 4 \\
\object{V384 Per} & \phantom{0}3 && 1.0 && 14.5 && 2.7 & 4 \\
\object{AFGL 3068} & 20 && 2.0 && 14.0 && 0.9 & 5 \\	
 & & & &&&& & \\
\object{TX Cam} & \phantom{0}7 && 1.0 && 18.5 && 3.0 & 4 \\
\object{GX Mon} & 20 && 0.5 && 18.7 && 0.1 & 4 \\
\object{WX Psc} & 40 && 0.4 && 19.3 && 1.0 & 4 \\	
\object{IK Tau}	& 10 && 0.3 && 19.0 && 0.4 & 4 \\
\object{IRC--10529}	& 16 && 0.2 && 14.0 && 3.8 & 4\\
\noalign{\smallskip}
\hline
\end{array}
$$
\end{table}

\subsubsection{Mass-loss rates from CO}
\label{s:MdotCO}
\begin{figure*}[t]
\raggedright 
{\includegraphics[width=18.5cm]{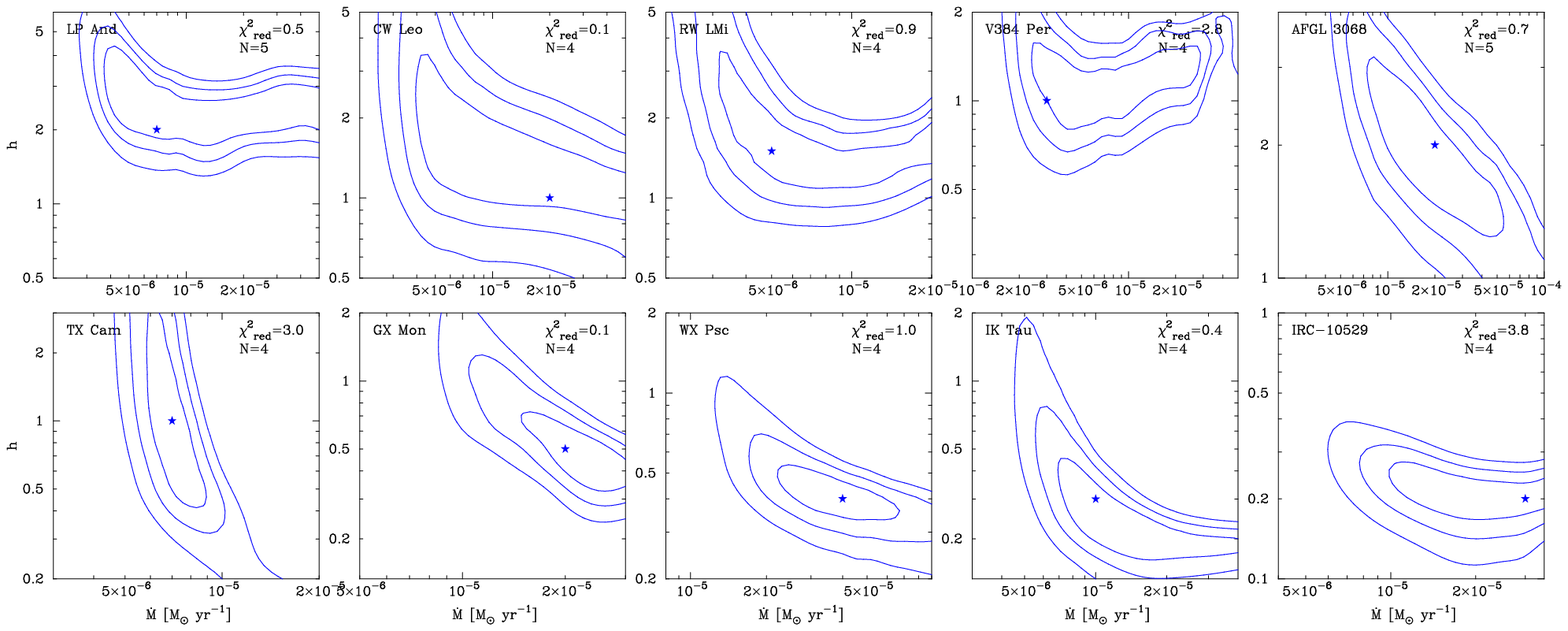}}
\caption{ $\mathrm{\chi^{2}}$-maps showing the level of consistency between the velocity-integrated CO line intensity from the observations and the line model when varying the mass-loss rate and the h-parameter. The innermost contour corresponds to a 68\% confidence level. The star indicates the best-fit model. The shapes of the line profiles were taken into account when choosing the best-fit model.}
\label{chi2-maps}
\vspace{0.2cm}
{\includegraphics[width=18.5cm]{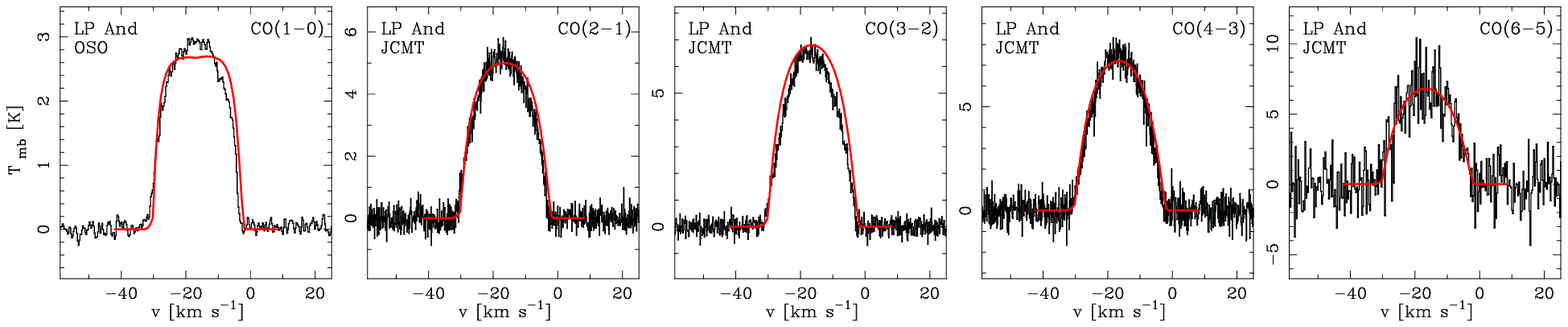}}
\caption{The observed CO $J$\,=\,1\,$\rightarrow$\,0 line from OSO and the 2\,$\rightarrow$\,1, 3\,$\rightarrow$\,2, 4\,$\rightarrow$\,3, and 6\,$\rightarrow$\,5 lines from JCMT (histogram) of the carbon star LP And, overlayed with the best-fit model ($\dot{M}$=7$\,\times\,$10$^{-6}$\,M$_{\odot}$\,yr$^{-1}$) lines (solid line) from the radiative transfer modelling. The reduced $\chi^{2}$ for this model is 0.25. See Sect.~\ref{sss:lpand}.}
\label{lpand_mod}
\vspace{0.2cm}
{\includegraphics[width=15.0cm]{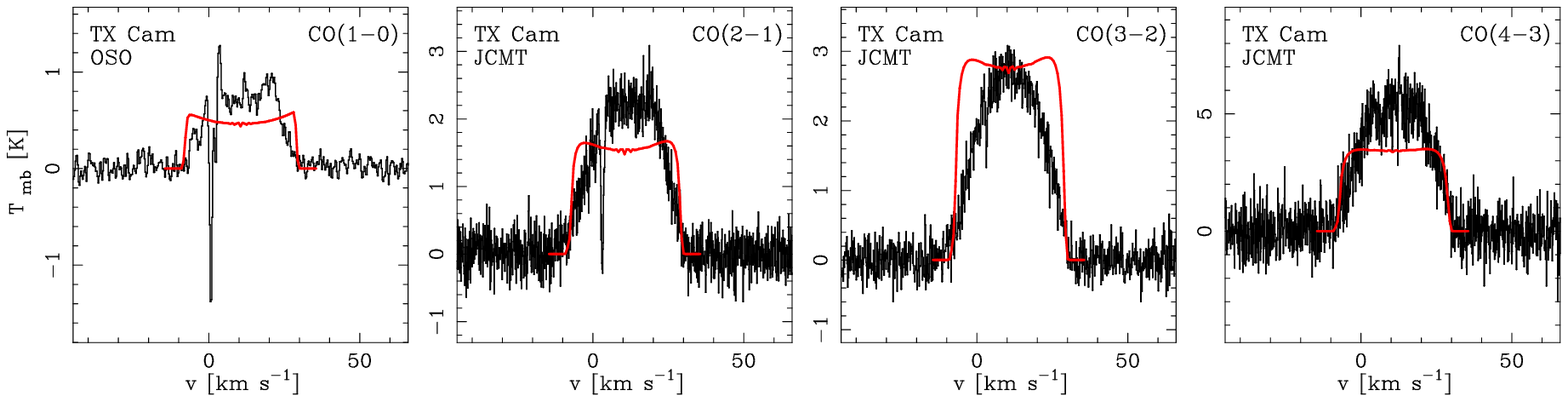}}
\caption{The observed CO $J$\,=\,1\,$\rightarrow$\,0 line from OSO and the 2\,$\rightarrow$\,1, 3\,$\rightarrow$\,2, and 4\,$\rightarrow$\,3 lines from JCMT (histogram) of the M-type star TX Cam, overlayed with the best-fit model ($\dot{M}$=7$\,\times\,$10$^{-6}$\,M$_{\odot}$\,yr$^{-1}$) lines (solid line) from the radiative transfer modelling. The reduced $\chi^{2}$ for this model is 3.0. See Sect.~\ref{sss:txcam}.}
\label{txcam_mod}
\end{figure*}

We are able to successfully model the CO line emission from all stars in our sample, both carbon stars and M-type stars, using a constant mass-loss rate. (The $J$=6$\rightarrow$5 line of WX Psc poses a problem that will be further discussed in Sect.~\ref{sss:wxpsc}.) The results are presented in Table~\ref{mdotCO}.  
Fig.~\ref{chi2-maps} shows $\chi^{\mathrm{2}}$-maps for all our sample stars. The difficulty in setting an upper limit on the mass-loss rate for high-mass-loss-rate stars using CO radio line intensities only is apparent. The saturation of the optically thick CO lines weakens significantly the dependence of the line intensity on the amount of CO in the envelope. 

In some cases, better constraints on the mass-loss rate can be set by including the shape of the line profile. Changing the mass-loss rate results in a changing photodissociation radius [see, e.g., model by \cite{stanetal95}], and this may result in a change of the line profile. Visual inspection of the line profiles (within the range of the innermost contour of the $\chi^{2}$-maps) was used to pick the best model presented in Table~\ref{mdotCO}, and when determining the error bars in Fig.~\ref{mdot}. In principle, it is possible to account for the shape of the line profile in a more quantitative way, i.e. by dividing the line into several velocity bins and calculating an error using $\chi^{2}$-statistics. However, this is problematic since the resulting $\chi^{2}$-values become very sensitive to very small variations in the line-profile shape which may not be possible to account for by changing the free parameters in the adopted model. Also, in cases like TX Cam (Fig.~\ref{txcam_mod}), a fit to the line-profile shape can never be found within the adopted model, either the adopted distance or the photodissociation radius must be changed (see Sect.~\ref{sss:txcam} for a discussion of this).

Figs~\ref{lpand_mod} and \ref{txcam_mod} give the comparisons between the observed and the best-fit model CO lines for the carbon star LP And and the M-type star TX Cam, respectively. These results are discussed in Sects~\ref{sss:lpand} and \ref{sss:txcam}, respectively. The results for the remaining stars are shown in Appendix~\ref{a:CO}. The differences between the integrated line intensities in the best-fit model and the observed lines are presented in Table~\ref{intensities_diff}.

\subsection{Dust emission modelling}
\label{s:res:dustmod}
\subsubsection{Strategy for finding the best-fit model}
\label{s:duststrat}

The observational constraints in the form of SEDs, covering the wavelength range 
$\sim$1$-$1300\,$\mu$m, are analyzed using $\chi^2$-statistics. When combining the measured fluxes into an SED, the variability of the star must be taken into account. This is particularly important for fluxes measured at shorter wavelengths, where the amplitude of the variation is larger. To assess how strongly the fluxes are affected, the mean and deviation was calculated from several measurements performed at different epochs. As a result a 60\% uncertainty was assumed for the J-band (1.24\,$\mu$m), 50\% for the H-band (1.63\,$\mu$m), 40\% for the K-band (2.19\,$\mu$m), and 30\% for the L- and M-bands (3.79 and 4.64\,$\mu$m, respectively). At wavelengths longer than 5\,$\mu$m, a calibration uncertainty of 20\% was added to 
the total error budget and it dominates the error (since there is limited information on the variability at these wavelengths). In some cases where there are many fluxes measured at roughly the same wavelength, we introduce an average value to avoid a bias in the $\chi^{2}$ calculation at certain wavelengths. (This is the reason why some of the fluxes presented in Table~\ref{fluxes} are not included in the analysis.) The SEDs constructed from the photometric flux densities are thus averages, where the variability of the star at different wavelengths is reflected in the error bars (typically larger uncertainties at shorter wavelengths).
The best-fit model is found from minimizing  
\begin{equation}
\label{chi2_sum}
\chi^2 = \sum^N_{i=1} \left [ \frac{(F_{\mathrm{mod},\lambda}-F_{\mathrm{obs},\lambda})}{\sigma_{\lambda}}\right ]^2, 
\end{equation} 
where $F_{\lambda}$ is the flux density and $\sigma_{\lambda}$ the uncertainty in the measured 
flux density at wavelength $\lambda$, and the summation is done over
all $N$ independent observations. 

\subsubsection{Results from dust emission modelling}

\begin{figure*}[t]
\raggedright 
{\includegraphics[width=18.2cm]{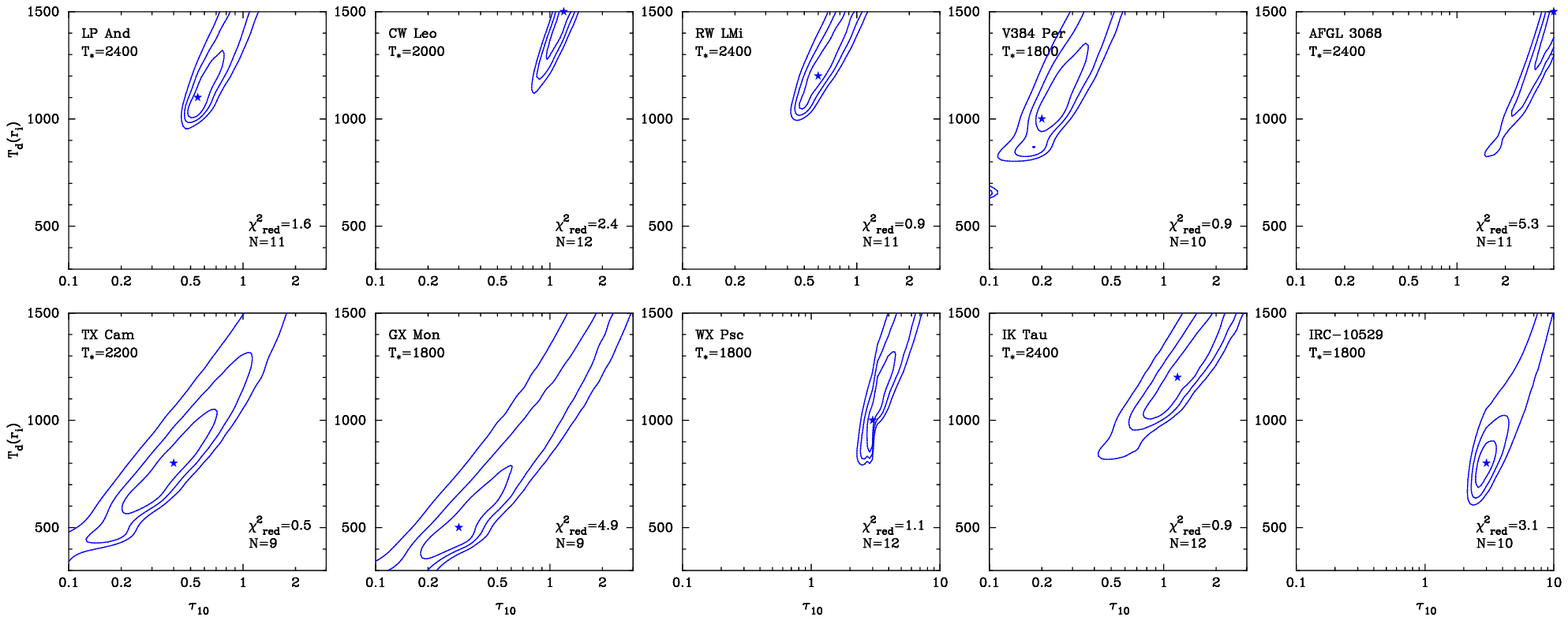}}
\caption{ $\mathrm{\chi^{2}}$-maps showing the level of consistency between the photometric fluxes and the dust-continuum model when varying $\tau_{10 \mu \mathrm{m}}$ and $T_{\mathrm{d}}$, at constant $T_{\mathrm{eff}}$. The innermost contour corresponds to a 68\% confidence level. The star indicates the best-fit model chosen using the strategy described in Sect.~\ref{s:duststrat}. All grids are terminated at $T_{\rm{d}}(r_{\rm{i}})$\,=\,1500\,K since a dust temperature at the inner radius larger than this is unlikely. This determines the best-fit models in the cases of CW Leo and AFGL 3068.}
\label{chi2-maps2}
{\includegraphics[width=18.1cm]{sed_c.epsi}}

\vspace{0.5cm}
{\includegraphics[width=18.1cm]{sed_m.epsi}}

\vspace{0.6cm}
{\includegraphics[width=18.1cm]{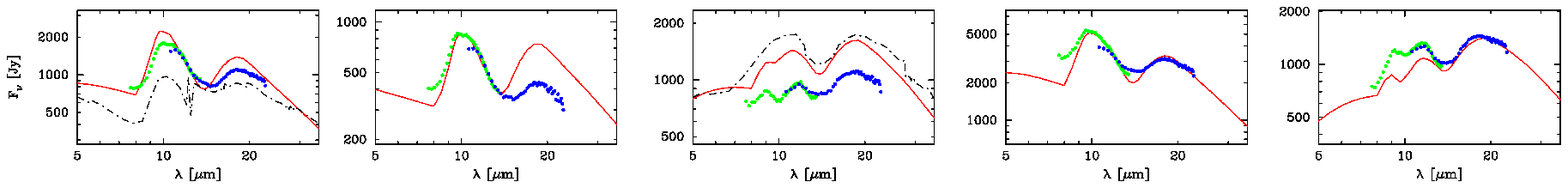}}
\caption{Observed photometric fluxes together with ISO-LWS and ISO-SWS spectra overlayed by the best-fit model SEDs derived from the dust radiative transfer model. The model is represented by the solid line in all panels. The dashed line shows the model SED when the interstellar extinction is not taken into account. The lower panels show the wavelength interval 5-35 $\mu$m in more detail for the M-type stars. The best-fit models are here shown together with the observed IRAS-LRS spectra (dotted line) and the ISO-SWS spectra (dot-dashed line) when available. The ISO-LWS spectra are plotted for all stars but GX Mon.}
\label{sed}
\end{figure*}

Table~\ref{dust} lists the results from the dust modelling, the dust temperature at the inner radius $T_{\mathrm{d}}(r_{\mathrm{i}})$, dust optical depth at 10\,$\mu$m $\tau_{10\mu \mathrm{m}}$, $\chi^{2}_{\mathrm{red}}$ from fitting the SED, and the number of observational constraints $N$.
$\chi^2$-maps, at constant best-fit stellar effective temperature, showing the level of consistency between the modelled SED and the photometric fluxes, when varying the dust temperature at the inner radius and the optical depth at 10\,$\mu \rm{m}$, are displayed in Fig.~\ref{chi2-maps2}. The SEDs from the dust-radiative-transfer model, with the observed photometric fluxes, are shown in Fig.~\ref{sed}.  $\chi^2_{\rm{red}}$ are generally close to 1 and the dust optical depth at $10\,\mu \rm{m}$ is rather well constrained for most sources (see Fig.~\ref{chi2-maps2}).

\begin{table}[h]
\caption{Results from the dust emission modelling.}
\label{dust}
$$
\begin{array}{p{0.25\linewidth}ccccccccc}
\hline
\noalign{\smallskip}
\multicolumn{1}{l}{{\mathrm{Source}}} &
\multicolumn{1}{c}{{T_{\star}}} &&
\multicolumn{1}{c}{{T_{\mathrm{d}}(r_{\mathrm{i}})}} &&
\multicolumn{1}{c}{{\tau_{10}}} &&
\multicolumn{1}{c}{{\mathrm{\chi^{2}_{red}}}} &&
\multicolumn{1}{c}{{N}} \\ 
 &
\multicolumn{1}{c}{[\mathrm{K}]} && \multicolumn{1}{c}{[\mathrm{K}]} && && &\\
\noalign{\smallskip}
\hline
\noalign{\smallskip}
\object{LP And} & 2400 && 1100 && 0.55 && 1.6 && 11 \\
\object{CW Leo} & 2000 && 1500 && 1.2\phantom{0} && 2.4 && 12 \\
\object{RW LMi} & 2400 && 1200 && 0.6\phantom{0} && 0.9 && 11 \\
\object{V384 Per} & 1800 && 1000 && 0.2\phantom{0} && 0.9 && 10 \\
\object{AFGL 3068} & 2400 && 1500 && 4.0\phantom{0}  && 5.3 && 11 \\	
 && && && && \\
\object{TX Cam}  & 2200 && 800 && 0.4\phantom{0} && 0.5 && \phantom{0}9 \\
\object{GX Mon}  & 1800 && 500 && 0.3\phantom{0} && 4.9 && \phantom{0}9 \\
\object{WX Psc} & 1800 && 1000 && 3.0\phantom{0} && 1.1 && 12 \\	
\object{IK Tau}  & 2400 && 1200 && 1.2\phantom{0}  && 0.9 && 12 \\
\object{IRC-10529} & 1800 && 800 && 3.0\phantom{0} && 3.1 && 10 \\

\noalign{\smallskip}
\hline
\end{array}
$$
\vspace{0.5cm}
\caption{Mass-loss rates and other quantities from the dynamical model.}
\label{mdotd}
$$
\begin{array}{p{0.18\linewidth}ccccc}
\hline
\noalign{\smallskip}
\multicolumn{1}{l}{{\mathrm{Source}}} &
\multicolumn{1}{c}{{\dot{M}}} &
\multicolumn{1}{c}{v_{\rm{dr}}} & 
\multicolumn{1}{c}{{\Psi}} &
\multicolumn{1}{c}{{\langle Q_{\mathrm{rp}} \rangle ^{\dagger}}} &
\multicolumn{1}{c}{{\dot{M}_{\rm{dyn}}/\dot{M}_{\rm{CO}}}}\\
&
\multicolumn{1}{c}{[\mathrm{10^{-6}\ M_{\odot}\ yr^{-1}}]} & 
\multicolumn{1}{c}{[\mathrm{km\,s^{-1}}]} &
\multicolumn{1}{c}{[\mathrm{10^{-3}}]} & &\\
\noalign{\smallskip}
\hline
\noalign{\smallskip}
\object{LP And} &  17 &  3.1  &  \phantom{0}1 &  0.09 &  2.5 \\
\object{CW Leo} & 21 & 1.9 &  \phantom{0}2 &  0.05 &  1.1 \\
\object{RW LMi} &  17 &  1.3 &  \phantom{0}2 &  0.08 &  3.3 \\
\object{V384 Per} &  \phantom{0}5 &  6.5 &  \phantom{0}2 &  0.11 &  1.6 \\
\object{AFGL 3068} &  62 &  0.7 &  \phantom{0}3 &  0.02 &  3.1 \\	
 & & & & & \\
\object{TX Cam}  &  \phantom{0}3 &  10.6 &  \phantom{0}4 &  0.11 &  0.5 \\
\object{GX Mon}  & \phantom{0}2 & 12.6 &  10 &  0.09 &  0.1 \\
\object{WX Psc} &  18 &  4.3 &  \phantom{0}3 &  0.06 &  0.5 \\	
\object{IK Tau}  &  \phantom{0}7 &  7.3 &  \phantom{0}2 &  0.10 &  0.7 \\
\object{IRC-10529} &  26 &  2.9 &  \phantom{0}2 &  0.06 & 1.6 \\
\noalign{\smallskip}
\hline
\end{array}
$$

\end{table}
%


\subsubsection{IR spectra -- a consistency check}
\label{ss:lrs}

Low resolution spectra in the wavelength range 7.5--23\,$\mu$m covering the silicate features at around 10 and 20\,$\mu$m, were measured with the Low Resolution Spectrograph (LRS) as part of the IRAS mission. Later, the Short Wavelength Spectrometer (SWS) and the Long Wavelength Spectrometer (LWS) onboard ISO measured medium and high resolution spectra in the wavelength range 2.38--45.2\,$\mu$m, and 43--197\,$\mu$m, respectively. SWS spectra also cover the silicate features and LWS spectra show the slope of the SED at longer wavelengths and is particularly sensitive to the cold dust.

The strengths of the silicate features are very sensitive to the dust optical depth [see for instance Fig.~1 in \citet{loon06}] and in turn the dust mass-loss rate. As a consistency check, the IRAS-LRS spectra and SWS spectra are plotted for the M-type stars in the lowest panel of Fig.~\ref{sed}. The carbon stars all have the SiC feature at 11 $\mu$m \citep[e.g.,][]{specetal97}. However, since SiC grains compose a small fraction of the total dust content ($\sim$10\%), and have a minor effect on the overall SED except for the 11 $\mu$m feature  \citep{suh00}, we have not included SiC in the dust model and consequently the observed spectra for the carbon stars are not shown. LWS spectra are available for all stars but one, and they are also plotted in Fig.~\ref{sed}.

A fit to the LRS and SWS spectra has not been attempted because of the difficulty in including them in a proper best-fit analysis. One complication is that the spectral shape and strength of the 9.7 $\mu$m silicate feature can change with the pulsational phase in stars with clear features and definite bolometric variability \citep{monnetal98}. This is likely valid for all M-type stars in our sample. Also, since these spectra are reasonably well fitted (see Fig~\ref{sed}), we expect the change in the resulting dust optical depth and inner radius, if the spectra were fitted in more detail (such as silicate features), to be small. A small change in the optical depth will result in an even smaller change in the estimated mass-loss rate from the dynamical model (see Table~\ref{dynerr}).

For some stars (CW Leo, RW LMi, AFGL 3068, and IRC-10529) there is a slight discrepancy between the slope of the LWS spectrum and that of the best-fit model. To perfectly reproduce the slope of the LWS spectra and the photometric fluxes at longer wavelengths, a varying mass-loss rate of these stars would probably have to be invoked. This has already been suggested for CW Leo by \citet{groe97}, who found that the dust density far from the star was underestimated by the best-fit $r^{-2}$ model. Other possible causes could be a difference between the optical properties of the model dust and the actual dust surrounding the star [see e.g. Figs~1 and 6 in \citet{suh00}], or the uncertainty in the long-wavelength dependence of the dust opacity. The fluxes beyond $\sim$\,800\,$\mu$m might also be affected by other processes than the dust emission, e.g., free-free emission and molecular line emission \citep{groe97,dehaetal07}. Bearing in mind the uncertainties in the circumstellar model (Sect.~\ref{s:circum}), and our aim to analyze all stars with the same assumptions we find that the LRS, SWS, and LWS data are well reproduced by our best-fit models.


\subsubsection{Mass-loss-rates from dust continuum and dynamical model}
\label{s:Mdotdust}
In the dynamical model, the density structure is calculated from the conservation of mass and the model is iterated until this gives an optical depth that agrees with the dust continuum optical depth obtained from the SED modelling (within one percent) and the resulting terminal gas expansion velocity from solving Eqns. (\ref{v_dust}) and (\ref{v_gas}) corresponds to the observed expansion velocity within 0.1\%.

Table~\ref{mdotd} lists mass-loss rates, $v_{\mathrm{dr}}$, $\Psi$, and $\langle Q_{\mathrm{rp}} \rangle^{\dagger}$ for the best-fit dynamical model. 


\section{Individual objects}
\label{s:ind}
All sources in the sample are well-studied both at IR and radio wavelengths. As established already in Sects~\ref{s:res:COmod} and \ref{s:res:dustmod}, we are able to reasonably successfully model the CO line emission, the integrated intensities and the line shapes, and the SEDs for all 10 stars using a constant mass-loss rate.

In the following discussion it should be kept in mind that the different methods measure the mass-loss rate at different epochs and over different time intervals. The size of the CO line emitting region decreases with the $J$-number and hence the different lines measure the mass-loss rate at slightly different epochs. The same is true for the SEDs where the emission at different wavelenghts (apart from the photospheric contribution) comes from different parts of the CSE. The estimated mass-loss rate should be considered as an average and it is a complicated synthesis of the different epochs and time scales.

\subsection{Carbon stars}

\subsubsection{LP And}
\label{sss:lpand}
The derived mass-loss rate from the CO modelling is 7$\,\times\,$10$^{-6}$\,M$_{\odot}$\,yr$^{-1}$, while the dynamical model gives a mass-loss rate of 1.7$\,\times\,$10$^{-5}$\,M$_{\odot}$\,yr$^{-1}$, a factor of 2.5 larger. The model lines agree very well with all five observed lines, $\chi^{2}_{\mathrm{red}}$=0.25 (Fig.~\ref{lpand_mod}). Previous mass-loss-rate estimates from CO radio lines have resulted in mass-loss rates close to our results, 1.2$\,\times\,$10$^{-5}$\,M$_{\odot}$\,yr$^{-1}$ (CO/H$_{2}$=1.1$\,\times\,$10$^{-3}$) \citep{teysetal06}, and 1.5$\,\times\,$10$^{-5}$\,M$_{\odot}$\,yr$^{-1}$ \citep{schoolof01}, both assuming a distance of 630\,pc. \citet{mensetal06} estimated the mass-loss rate to be 1.9$\,\times\,$10$^{-5}$\,M$_{\odot}$\,yr$^{-1}$ from a dust radiative transfer model, assuming a distance of 740\,pc and a dust-to-gas mass ratio of 0.0039. The corresponding mass-loss rate at the distance used and the dust-to-gas ratio obtained in our model, would be 1.1$\,\times\,$10$^{-5}$\,M$_{\odot}$\,yr$^{-1}$, in good agreement with our result.   
 
\citet{nerietal98} found a slight asymmetry in the CSE from maps of the CO($J$\,=\,1\,$\rightarrow$\,0) emission. \citet{maurhugg06}, using optical images in scattered light, suggested that the envelope around LP And may have an interesting structure, but concluded that the signal-to-noise in their data is too low to make any firm statement. 

\subsubsection{CW Leo (IRC+10216)}
\label{sss:cwleo}
Being the closest carbon star, CW Leo is certainly the most well-studied. We estimate the mass-loss rate to be 2$\,\times\,$10$^{-5}$\,M$_{\odot}$\,yr$^{-1}$ from the CO line modelling and 2.1$\,\times\,$10$^{-5}$\,M$_{\odot}$\,yr$^{-1}$ from the dynamical model.
Previous estimates of its mass-loss rate from CO is 1.2$\,\times\,$10$^{-5}$\,M$_{\odot}$\,yr$^{-1}$ \citep{teysetal06}, and 1.5$\,\times\,$10$^{-5}$\,M$_{\odot}$\,yr$^{-1}$ \citep{schoolof01}, both assuming a distance of 120\,pc. \citet{groeetal98} did a very thorough investigation of the CO radio line emission toward CW Leo and estimated the present day mass-loss rate to (1.5$\pm$0.3)$\,\times\,$10$^{-5}$\,M$_{\odot}$\,yr$^{-1}$ and the distance to lie between 110 and 135\,pc. 

Several reports on structures in the circumstellar environment around CW Leo can be found in the literature \citep[e.g., ][]{maurhugg99,fongetal06}. Optical images show arc-like structures on arcsecond scales in the CSE, but whether this is due to changes in the mass-loss rate over time, is still under debate [e.g., \citet{leaoetal06} and references therein].  

\subsubsection{V384 Per}
\label{sss:v384per}
The best-fit CO model gives a mass-loss rate of 3$\,\times\,$10$^{-6}$\,M$_{\odot}$\,yr$^{-1}$. However, as can be seen in Fig.~\ref{v384per_mod}, the model $J$\,=\,1\,$\rightarrow$\,0, 2\,$\rightarrow$\,1, and 4\,$\rightarrow$\,3 lines are all 15-20\% weaker than the observed lines, while the model $J$\,=\,3$\rightarrow$\,2 line is 65\% stronger than the observed line. This results in a rather high $\chi^{2}_{\rm{red}}$=2.8. The most likely explanation is a calibration error in the $J$\,=\,3\,$\rightarrow$\,2 line. The mass-loss rate from the dynamical model is higher, $\dot{M}\,$=\,5$\,\times\,$10$^{-6}$\,M$_{\odot}$\,yr$^{-1}$. \citet{schoolof01} estimated $\dot{M}\,$=\,3.5$\,\times\,$10$^{-6}$\,M$_{\odot}$\,yr$^{-1}$ at 560\,pc from CO line modelling. 

\subsubsection{AFGL 3068 (LL Peg)}
\label{sss:afgl3068}
Our model of the CO line emission reproduces all five lines for a constant mass-loss rate of 2$\,\times\,$10$^{-5}$\,M$_{\odot}$\,yr$^{-1}$ with a $\chi^{2}_{\rm{red}}$=0.7. The dynamical model gives $\dot{M}\,$=\,6.2$\,\times\,$10$^{-5}$\,M$_{\odot}$\,yr$^{-1}$, slightly more than a factor of three higher. Earlier estimates of the mass-loss rate from CO line modelling have resulted in $\dot{M}\,$=\,6.0$\,\times\,$10$^{-5}$\,M$_{\odot}$\,yr$^{-1}$ \citep{teysetal06} at 1000\,pc, and 2.0$\,\times\,$10$^{-5}$\,M$_{\odot}$\,yr$^{-1}$ \citep{woodetal03} at 820\,pc. \citet{teysetal06} deduced a much lower temperature in the outer envelope than we calculate in our model and this is the reason for the discrepancy between the estimates. 

There is no doubt that the CSE around this source is asymmetric and have density structures that cannot be reconciled with a spherically symmetric CSE \citep{maurhugg06}. Radial CO ($J$\,=\,1\,$\rightarrow$\,0) visibility profiles show a compact inner envelope surrounded by a large outer shell. The outer shell is strongly bipolar and there are even indications that it might be detached \citep{nerietal98}.

\subsubsection{RW LMi (CIT 6)}
\label{sss:rwlmi}
We find the CO line data to be consistent with a mass-loss rate of 5$\,\times\,$10$^{-6}$\,M$_{\odot}$\,yr$^{-1}$. From the dynamical model the mass-loss rate is estimated to be 17$\,\times\,$10$^{-6}$\,M$_{\odot}$\,yr$^{-1}$. 
\citet{teysetal06} found it necessary to infer a change in the mass-loss rate about 250 years ago to fit the CO intensity measured at offset positions; the mass-loss rate has decreased from 7.5$\,\times\,$10$^{-6}$\,M$_{\odot}$\,yr$^{-1}$ to 5.0$\,\times\,$10$^{-6}$\,M$_{\odot}$\,yr$^{-1}$. The two component structure was already noticed by \citet{nerietal98} when first inspecting the interferometric data. \citet{schoolof01} estimated $\dot{M}\,$=\,6.0$\,\times\,$10$^{-6}$\,M$_{\odot}$\,yr$^{-1}$ from CO line modelling using the same distance as \citet{teysetal06}, 440 pc.

\subsection{M-type stars}

\subsubsection{TX Cam}
\label{sss:txcam}
The best-fit CO model gives $\dot{M}$\,=\,7$\,\times\,$10$^{-6}$\,M$_{\odot}$\,yr$^{-1}$ [the same rate was derived by \citet{delgetal03}], but the model does not reproduce the shape of the observed line profiles (Fig~\ref{txcam_mod}). The observed $J$\,=\,1\,$\rightarrow$\,0 and 2\,$\rightarrow$\,1 lines are flat-topped, which suggests unresolved, optically thin emission, while the model gives double-peaked lines, which suggests resolved, optically thin emission. The observed $J$\,=\,3\,$\rightarrow$\,2 and 4\,$\rightarrow$\,3 lines are parabolic and thus optically thick, while the model produces an optically thin, resolved $J$\,=\,3\,$\rightarrow$\,2 line and a flat-topped 4\,$\rightarrow$\,3 line. The simplest explanation would be that the distance is underestimated. A larger distance requires a higher mass-loss rate to reproduce the line intensities and hence gives line profiles that better reproduce the observed ones. An increased distance also decreases the angular size, but this is partly counteracted by the fact that the CO photodissociation radius increases with the mass-loss rate. Another explanation could be that the size of the CO envelope in this object is smaller than the photodissociation radius given by the model of \citet{mamoetal88}. Decreasing the size of the CO envelope solves the problem with the model producing resolved lines. Also, a smaller radius will require a higher mass-loss rate to reproduce the observed intensities. This also increases the optical depth. However, the dynamical model estimate of the mass-loss rate is 3$\,\times\,$10$^{-6}$\,M$_{\odot}$\,yr$^{-1}$, a little more than a factor of 2 less than the CO modelling.

\subsubsection{GX Mon}
\label{sss:gxmon}
The dynamical model gives a mass-loss rate estimate (2$\,\times\,$10$^{-6}$\,M$_{\odot}$\,yr$^{-1}$) one order of magnitude lower than the CO model (2$\,\times\,$10$^{-5}$\,M$_{\odot}$\,yr$^{-1}$). The best-fit SED model found according to the strategy described above gives a dust temperature at the inner radius of 500\,K. Fig. \ref{chi2-maps2} shows the $\chi^{2}$-maps from the SED-fitting and in the case of GX Mon, the model is not very well constrained. The major dust condensation phase in M-type stars is thought to occur at temperatures between 1200\,K and 800\,K \citep{whit03}. With the low dust optical depth derived for GX Mon a high dust-to-gas ratio ($\Psi=0.01$) is needed to accelerate the gas to the measured expansion velocity. [Setting $T_{\rm{d}}(r_{\rm{i}})$\,=\,1000\,K, requires a higher optical depth to fit the model to the observed fluxes ($\tau_{10}$=0.85), and results in $\dot{M}\,$=\,6.3$\,\times\,$10$^{-6}$\,M$_{\odot}$\,yr$^{-1}$ from the dynamical model, somewhat more in line with the CO results. The measured expansion velocity would be reached with $\Psi\approx8.3\times10^{-3}$.] A possible explanation for the low temperature at the inner radius is that the star has undergone a rather drastic decrease in the mass-loss rate recently. The dust envelope from an earlier epoch would have moved outward and cooled to the temperature obtained from the modelling, and the photometric fluxes would thus probe this 'older' shell. In such a case the dynamical model would produce an erroneous result.
 
\subsubsection{WX Psc}
\label{sss:wxpsc} 
The $J$\,=\,1\,$\rightarrow$\,0, 2\,$\rightarrow$\,1, 3\,$\rightarrow$\,2, and 4\,$\rightarrow$\,3 lines are well-fitted by a constant mass-loss rate of $\dot{M}$\,=\,4$\,\times\,$10$^{-5}$\,M$_{\odot}$\,yr$^{-1}$. We are, however, not able to fit the $J$\,=\,6\,$\rightarrow$\,5 line using this model. The problem of not being able to model higher-frequency CO lines from this source using a constant mass-loss rate has been previously noted by \citet{kempetal03}, \citet{teysetal06} and \citet{decietal07}. The difficulty of getting well-calibrated data for these higher frequency lines should be kept in mind when drawing conclusions based on this kind of data. As an example, the integrated intensity in the 6\,$\rightarrow$\,5 line in this work is a factor of two stronger than that reported by \citet{kempetal03}. A model fitting only our 6\,$\rightarrow$\,5 line results in $\dot{M}$\,=\,4$\,\times\,$10$^{-6}$\,M$_{\odot}$\,yr$^{-1}$. \citet{decietal07} introduce a rather complicated mass-loss history, and estimate the mass-loss rate to be 6$\,\times\,$10$^{-6}$\,M$_{\odot}$\,yr$^{-1}$ in the region where the 6\,$\rightarrow$\,5 line is mainly coming from. Higher transition CO lines are within the range of archived ISO LWS spectra (43 -- 197 $\mu$m). The CO $J$\,=\,16\,$\rightarrow$\,15 and $J$\,=\,18$\rightarrow$\,17 lines were modelled using our best-fit model. The resulting peak intensities [4.6\,$\times$\,10$^{-22}$\,W\,cm$^{-2}$\,(km/s)$^{-1}$ and 5.0\,$\times$\,10$^{-22}$\,W\,cm$^{-2}$\,(km/s)$^{-1}$, respectively] would result in a S/N\,$\approx$\,5 [the resolution-corrected $1\sigma$ noise level is 8.9\,$\times$\,10$^{-23}$\,W\,cm$^{-2}$\,(km/s)$^{-1}$]. None of the CO lines within the range of the ISO LWS were detected, and this is in line with the weakness of the observed $J$\,=\,6\,$\rightarrow$\,5 line. Thus, the results reported by us and previous authors suggest a recent decrease in the mass-loss rate in this source. From the dynamical model we find $\dot{M}$=1.8$\,\times\,$10$^{-5}$\,M$_{\odot}$\,yr$^{-1}$.

\subsubsection{IK Tau}
\label{sss:iktau}
Modelling IK Tau with a constant mass-loss rate of 1$\,\times\,$10$^{-5}$\,M$_{\odot}$\,yr$^{-1}$ gives a good fit to the four available CO lines (Fig.~\ref{iktau_mod}). This rate is a factor of two higher than that found by \citet{teysetal06} at a somewhat smaller distance of 250\,pc. The dynamical model gives $\dot{M}$=7$\,\times\,$10$^{-6}$\,M$_{\odot}$\,yr$^{-1}$. 

\subsubsection{IRC-10529}
\label{sss:irc10529}
When modelling IRC-10529, we encounter a similar problem as for V384 Per. We find that the best-fit CO model with $\dot{M}$\,=\,3$\,\times\,$10$^{-5}$\,M$_{\odot}$\,yr$^{-1}$ has a $\chi^{2}_{\rm{red}}$=3.8. The $J$\,=\,1\,$\rightarrow$\,0 and 3\,$\rightarrow$\,2 lines are fitted within the observational uncertainties, but the model 2\,$\rightarrow$\,1 line is 45\% too weak, and the 4\,$\rightarrow$\,3 line is 40\% too strong (Fig~\ref{irc10529_mod}). We do not find it likely that this is due to modulations in the mass-loss rate, since the emitting regions for the 1\,$\rightarrow$\,0 and 2\,$\rightarrow$\,1 lines overlap to a large extent. It is more likely due to observational problems. The dynamical model gives $\dot{M}$\,=\,2.6$\,\times\,$10$^{-5}$\,M$_{\odot}$\,yr$^{-1}$, in good agreement with the estimate from the CO model.  


\section{Discussion}
\label{s:discuss}

\subsection{Dependence on parameters}
\label{ss:dep}

The model CO line intensities dependence on the $h$-parameter and $\dot{M}$ is displayed in Fig.~\ref{chi2-maps}. A further discussion on how the resulting line intensities depend on various input parameters can be found in \citet{schoolof01}.

To evaluate the sensitivity of the results of the dynamical model to the different derived or assumed input parameters, we varied the parameters in Eqns~(\ref{v_dust}) and (\ref{v_gas}) by $\pm$\,50\%. Shown in Table~\ref{dynerr} is the needed change of the dust-to-gas ratio, $\Psi$, and the mass-loss rate, $\dot{M}$, to reach the observed expansion velocity and derived optical depth for RW LMi. As can be seen, the dynamical model results are strongly dependent on the assumed dust parameters, which are not well known. 

A trend with mass-loss rate can be seen in the dynamical model. For TX Cam, an intermediate-mass-loss-rate object ($\dot{M}$\,=\,3$\,\times\,$10$^{-6}$\,M$_{\odot}$\,yr$^{-1}$ from the dynamical model), an increase in the mass-loss rate by 50\% leads to an increased expansion velocity by 6\%. For RW LMi, a high-mass-loss-rate object ($\dot{M}$\,=\,1.7$\,\times\,$10$^{-5}$\,M$_{\odot}$\,yr$^{-1}$) the same increase in the mass-loss rate will lead to a decrease in the expansion velocity by 15\%. This is due to the fact that a higher mass-loss rate leads to a higher dust optical depth resulting in a radiation field that decreases more rapidly with radius and with it the radiation pressure efficiency. Nevertheless, this does not lead to a problem finding the correct mass-loss rate. In some cases, two mass-loss rates give the same expansion velocity, but there is only one solution that produces the correct dust optical depth. 

We have decided not to display any error bars on the estimates from the dynamical model in Fig.~\ref{mdot}. The reason for this is that the resulting (formal) error bars given by the dust radiative transfer model (variations in inner radius and optical depth, Fig.~\ref{chi2-maps2}) are small, and ignores the substantial uncertainties in the adopted dust parameters.

\begin{table}
\caption{The resulting change in mass-loss rate ($\dot{M}$) and dust-to-gas ratio ($\Psi$) when varying the input parameters in the dynamical model of RW LMi.}
\label{dynerr}
$$
\begin{array}{p{0.1\linewidth}cp{0.2\linewidth}ccc}
\hline
\noalign{\smallskip}
\multicolumn{1}{l}{{\mathrm{Parameter}}} &&
\multicolumn{1}{l}{{\rm{Change}}} &&
\multicolumn{1}{c}{{\mathrm{\dot{M}}}} &
\multicolumn{1}{c}{{\mathrm{\Psi}}} \\
\noalign{\smallskip}
\hline
\noalign{\smallskip}
$L$ && +50\% && +\phantom{1}50\% & -\phantom{1}35\% \\
        && $-$50\% && -\phantom{1}50\% & +100\% \\
$M_{*}$ && +50\% && -\phantom{1}25\% & +\phantom{1}30\% \\
        && $-$50\% && +\phantom{1}30\% & -\phantom{1}30\% \\	
$\sigma_{\rm{d}}$ && +50\% && +\phantom{1}50\% & -\phantom{1}35\% \\
        && $-$50\% && -\phantom{1}50\% & +125\% \\
$m_{\rm{d}}$ && +50\% && -\phantom{1}35\% & +\phantom{1}60\% \\
        && $-$50\% && +105\% & -\phantom{1}55\% \\
$\tau_{10}$ && +50\% && +\phantom{1}35\% & +\phantom{1}10\% \\
        && $-$50\% && -\phantom{1}40\% & -\phantom{1}10\% \\
{\bf $Q_{\lambda}$} && +50\% && +\phantom{1}50\% & -\phantom{1}35\% \\
        && $-$50\% && -\phantom{1}50\% & +100\% \\
\noalign{\smallskip}
\hline
\end{array}
$$
\end{table}

\subsection{Comparing the two methods}
\label{ss:comp}
It is very difficult to assess which method is the most reliable for estimating mass-loss rates, since both are dependent on a number of assumptions, which are more or less well-founded. For instance, the dynamical model assumes an entirely dust driven wind, while recent results suggest that there might be other factors at play \citep{woit06,hofnande07}. 

\begin{figure}
\centering{   
\includegraphics[width=10cm]{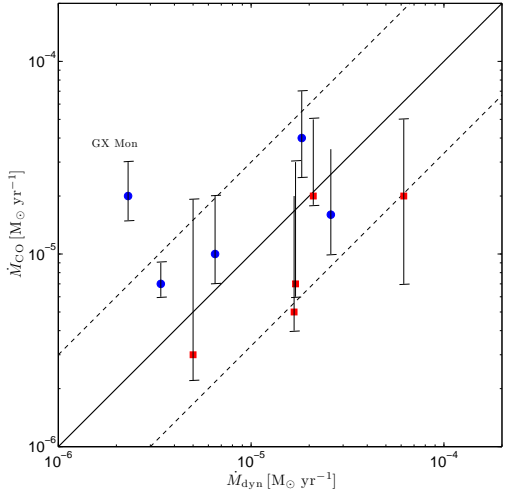}
\caption{Comparing the mass-loss-rate results from the CO model and the dynamical model. Dots mark the M-type stars, and squares the carbon stars. The full drawn line represents a one-to-one relation and the dashed lines subtend the range within a factor of 3.}
\label{mdot}}
\end{figure}

In all cases but GX Mon, for which we have good reasons to believe that the dynamical model gives an erroneous result (see Sect.~\ref{sss:gxmon} for an explanation), the mass-loss rates estimated from the dynamical model and the CO line modelling agree within a factor of $\sim$\,3. A comparison is shown in Fig.~\ref{mdot}, where the full drawn line marks the one-to-one relation, while the dashed lines subtend the range within a factor of 3. [Error bars represents the range within the innermost contour in the $\chi^{2}$-maps from the CO model. Here the shape of the line profiles have also been taken into account.] Since the two methods give independent estimates of the mass-loss rate, this result can be used as a guideline to the uncertainty in the mass-loss-rate estimates within the adopted circumstellar model. \citet{herahony05} found in their comparison of mass-loss rates derived from ISO-SWS and CO mass-loss rates taken from the literature for low-mass-loss-rate M-type stars that the estimated mass-loss rates can differ up to a factor of ten.

In principle, it is possible to check whether the dust results obtained from the dynamical model ($v_{\rm{dr}}$, $\Psi$, $\langle Q_{\mathrm{rp}} \rangle^{\dagger}$) are consistent with those used in the CO radiative transfer modelling ($h$, $\langle Q_{\mathrm{rp}} \rangle^{\dagger}$). However, we do not expect to get a perfect agreement since in the CO radiative transfer modelling the $h$-parameter is adjusted until the radial kinetic temperature profile results in reasonable relative strengths of the CO lines. Hence, it is affected by possible missing heating and cooling terms in the description of the thermodynamics of the circumstellar gas. However, as outlined in Sect.~\ref{s:COmod}, this does not affect the mass-loss-rate estimates.

For the carbon stars in our sample the dust-to-gas ratio, $\Psi$, is estimated to be $0.001<\Psi<0.003$. \citet{groeetal98} derived dust-to-gas ratios for a sample of 36 carbon stars and found a rather constant value of 0.0025 up to $P\approx500$ where a strong increase with period is found. This agrees rather well with our results. However for AFGL 3068 they find $\Psi\approx0.02$, in strong excess of our result. For the M-type stars we find somewhat larger dust-to-gas ratios, $0.002<\Psi<0.004$ (GX Mon excluded). \citet{justetal94} modelled three thick M-type CSEs and found $0.003<\Psi<0.006$.

The CO mass-loss-rate estimate is lower than the estimate from the dynamical model for all five carbon stars, see Fig.~\ref{mdot}. The opposite can be seen for the M-type stars, where the CO model gives a mass-loss-rate estimate higher than the dynamical model for three out of four M-type stars (GX Mon excluded). Keeping in mind that the sample presented here is too small to make a general statement, we suggest that a possible explanation to this is that the adopted CO/H$_{2}$-ratio is too low, on average, for the M-type stars and too high for the carbon stars. This is also in line with recent chemical models \citep{cher06}, where a non-equilibrium chemical model gives a CO-abundance of 6\,$\times$\,10$^{-4}$ in M-type stars (C/O=0.75) at 5 stellar radii. For the carbon stars (C/O=1.1), the resulting CO-abundance at the same distance from the star is found to be 9\,$\times$\,10$^{-4}$.  Adopting these abundances would require a higher mass-loss rate for the carbon stars and a lower mass-loss rate for the M-type stars to reproduce the observed CO line intensities.

\subsection{Comments on some results by Kemper et al. (2003)}

\begin{figure}
\centering{   
\includegraphics[width=7cm,height=7cm,angle=-90]{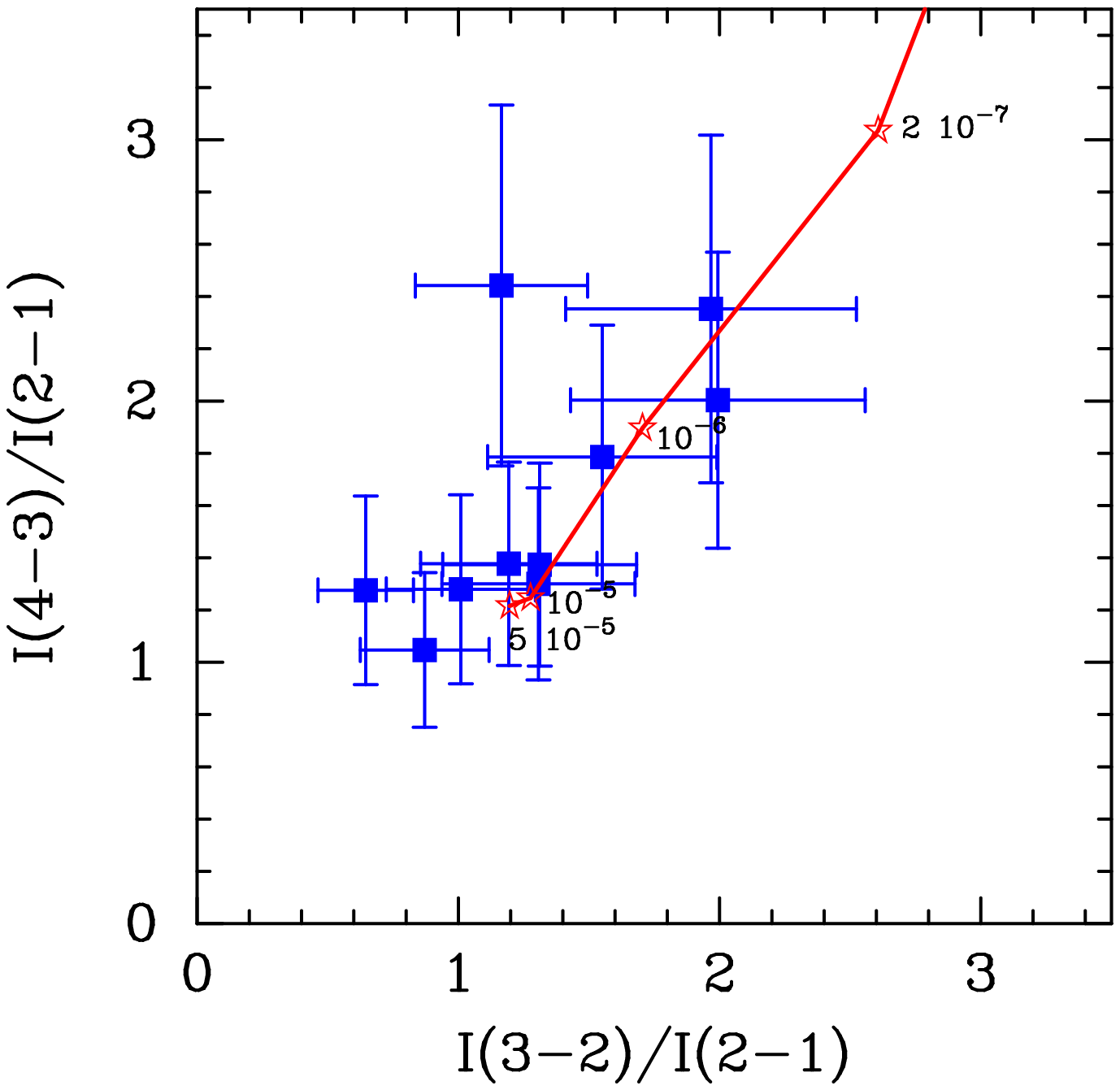}
\caption{Observed line intensity ratios for the sample sources (squares) compared with predictions from the standard model for various mass loss rates (stars).}
\label{ratios}}
\end{figure}

\citet{kempetal03} made non-LTE radiative transfer modelling of CO $J$\,=\,2\,$\rightarrow$\,1 to $J$\,=\,6\,$\rightarrow$\,5 emission toward a sample of AGB stars and red supergiants. In a line ratio diagram they compared modelled line intensity ratios, for standard CSEs with different mass-loss rates, to those found from observations (Fig.~4 in their paper). Kemper et al. found that they could not reproduce the observed line intensity ratios. On the contrary, the standard model described above (Sect.~\ref{s:COmod}), generally gives a very good fit to the observed line intensities as indicated by a $\chi_{\mathrm{red}}$ of the order of unity for the majority of the sample sources. This is also illustrated in Fig.~\ref{ratios} where line intensity ratios calculated from our standard model using $v_{\rm{e}}$\,=\,15\,km\,s$^{-1}$ and $h$\,=\,0.5 for a range of mass-loss rates are compared to the observations. The model CSE was placed at a distance of 1\,kpc. The interconnected stars are the results from the standard model and the squares show the observed line ratios.

\citet{kempetal03} also found that the magnitude of the stochastic velocity strongly affects the line intensities, since it determines the size of the region where the line is formed [$v_{\rm{sto}}=({v_{\rm{turb}}^{2}+v_{\rm{therm}}^{2}})^{0.5}\approx v_{\rm{turb}}$, since $v_{\rm{therm}}$ is the local thermal broadening and $v_{\rm{therm}}\ll v_{\rm{turb}}$]. We have performed the same test as in their Fig.~10, changing the turbulent velocity from 0.05 to 2\,km\,s$^{-1}$, to see how this affects the intensity of the CO($J$\,=\,3\,$\rightarrow$\,2) line. We used the same standard model as above with a mass-loss rate of $\dot{M}$=1\,$\times$\,10$^{-5}$\,M$_\odot$\,yr$^{-1}$. \citet{kempetal03} saw a decrease in intensity from around $T_{\rm{mb}}$=2.4\,K at $v_{\rm{sto}}$=0.05\,km\,s$^{-1}$ to $T_{\rm{mb}}$=0.2\,K at $v_{\rm{sto}}$=2.0\,km\,s$^{-1}$. We, however, see the opposite trend and a much weaker effect; $T_{\rm{mb}}$=1.05\,K at $v_{\rm{turb}}$=0.05\,km\,s$^{-1}$ to $T_{\rm{mb}}$=1.28\,K at $v_{\rm{turb}}$=2.0\,km\,s$^{-1}$. It is difficult to outline exactly how the line intensity is affected by the turbulent velocity, as it also affects the excitation of the molecules. However, to a first approximation, with an increased turbulent width the number of molecules per unit velocity interval decreases. On the other hand, as already mentioned above, it also increases the size of the interacting region. Under certain circumstances (spherically symmetric CSE, constant excitation temperature, $v_{\rm{turb}}\,\ll\,v_{\rm{e}}$) these two effects cancel out and the line intensity becomes independent of the turbulent velocity \citep[see for example][]{morris75,olofetal82}. 

In conclusion, we are not able to reproduce the results reported by \citet{kempetal03} using our model. We confirmed our result using two additional codes; an accelerated Monte Carlo code \citep{hogevdta00} and an Accelerated Lambda Iteration (ALI) code \citep{maeretal08}. The results from the excitation analysis were then combined with different ray tracers [the one used in this paper and the code described in \citet{hogevdta00}] to get the resulting line profiles. All results were in good agreement with ours.

\section{Conclusions}
\label{s:conc}
We have presented mass-loss-rate estimates, derived using two independent methods, for a sample of 10 intermediate- to high-mass-loss-rate, well-studied AGB stars, and arrive at the following conclusions:

\begin{itemize}
\item{The mass-loss rate estimates obtained using the two different methods agree within a factor of $\sim$\,3, suggesting that this is (at least) the uncertainty in present mass-loss-rate estimates, e.g., using CO radio line emission. The average of the ratio between the two mass-loss-rate estimates is $\langle\dot{M}_{\rm{dyn}}/\dot{M}_{\rm{CO}}\rangle=2.3\pm1.0$ for the carbon stars, and $\langle\dot{M}_{\rm{dyn}}/\dot{M}_{\rm{CO}}\rangle=0.8\pm0.5$ for the M-type stars (excluding GX Mon).}
\item{In most cases, it is not possible to set an upper limit to the mass-loss rate for the high-mass-loss-rate objects by modelling only the integrated CO line intensities. The analysis can be improved by taking the shape of the line profile into account.}
\item{We are able to successfully model the CO($J$\,=\,1\,$\rightarrow$\,0 to 4\,$\rightarrow$\,3) data of all stars using a constant mass-loss rate. However, in most cases there are various indications that our 'simple' circumstellar model does not fully reproduce the data. Effects of time-variable mass loss, and non-isotropic mass loss may be present.}
\item{We do not find any trends with mass-loss rate when comparing the two models, but we find a possible dependence on chemical type. Generally, the CO model estimates a lower mass-loss rate for the carbon stars than the dynamical model, while the opposite behaviour can be observed for the M-type stars. We conclude that a CO/H$_{2}$-ratio increased and decreased by about a factor of 1.5 for the M- and C-stars, respectively, would give a very good agreement between the methods, but caution that this is a tentative conclusion considering that the sample is small, and there may be other systematic errors.}
\end{itemize}

\begin{acknowledgements}
The authors acknowledge support from the Swedish Research Council.
This article made use of data obtained through the JCMT archive as Guest User 
at the Canadian Astronomy Data Center, which is operated by the Dominion
Astrophysical Observatory for the National Research Council of Canada's
Herzberg Institute of Astrophysics. Finally the authors would like to thank Franz Kerschbaum for providing us with a compilation of near-IR data needed for the dust radiative transfer modelling. 
\end{acknowledgements}

\bibliographystyle{aa}
\bibliography{HighMdot}

\begin{thebibliography}{94}
\expandafter\ifx\csname natexlab\endcsname\relax\def\natexlab#1{#1}\fi

\bibitem[{{Bagnulo}(1996)}]{bagnulo96}
{Bagnulo}, S. 1996, PhD thesis, Ph.D.~dissertation, Queen's University Belfast,
  N.~Ireland, (1996)

\bibitem[{{Bains} {et~al.}(2003){Bains}, {Cohen}, {Louridas}, {Richards},
  {Rosa-Gonz{\'a}lez}, \& {Yates}}]{bainetal03}
{Bains}, I., {Cohen}, R.~J., {Louridas}, A., {et~al.} 2003, \mnras, 342, 8

\bibitem[{{Battinelli} \& {Demers}(2004)}]{battdeme04}
{Battinelli}, P. \& {Demers}, S. 2004, \aap, 418, 33

\bibitem[{{Beichman} {et~al.}(1988){Beichman}, {Neugebauer}, {Habing}, {Clegg},
  \& {Chester}}]{iras88}
{Beichman}, C.~A., {Neugebauer}, G., {Habing}, H.~J., {Clegg}, P.~E., \&
  {Chester}, T.~J., eds. 1988, {Infrared astronomical satellite (IRAS) catalogs
  and atlases. Volume 1: Explanatory supplement}, Vol.~1

\bibitem[{{Bergeat} {et~al.}(2001){Bergeat}, {Knapik}, \&
  {Rutily}}]{bergetal01}
{Bergeat}, J., {Knapik}, A., \& {Rutily}, B. 2001, \aap, 369, 178

\bibitem[{{Bloecker}(1995)}]{bloe95}
{Bloecker}, T. 1995, \aap, 297, 727

\bibitem[{{Bowen}(1988)}]{bowe88}
{Bowen}, G.~H. 1988, \apj, 329, 299

\bibitem[{{Cherchneff}(2006)}]{cher06}
{Cherchneff}, I. 2006, \aap, 456, 1001

\bibitem[{{Cohen} {et~al.}(1992){Cohen}, {Walker}, \& {Witteborn}}]{coheetal92}
{Cohen}, M., {Walker}, R.~G., \& {Witteborn}, F.~C. 1992, \aj, 104, 2030

\bibitem[{{Decin} {et~al.}(2007){Decin}, {Hony}, {de Koter}, {Molenberghs},
  {Dehaes}, \& {Markwick-Kemper}}]{decietal07}
{Decin}, L., {Hony}, S., {de Koter}, A., {et~al.} 2007, ArXiv e-prints, 708

\bibitem[{{Dehaes} {et~al.}(2007){Dehaes}, {Groenewegen}, {Decin}, {Hony},
  {Raskin}, \& {Blommaert}}]{dehaetal07}
{Dehaes}, S., {Groenewegen}, M.~A.~T., {Decin}, L., {et~al.} 2007, \mnras, 377,
  931

\bibitem[{{Dorschner} \& {Henning}(1995)}]{dorshenn95}
{Dorschner}, J. \& {Henning}, T. 1995, \aapr, 6, 271

\bibitem[{{Epchtein} {et~al.}(1985){Epchtein}, {Matsuura}, {Braz}, {Lepine},
  {Picazzio}, {Marques Dos Santos}, {Boscolo}, {Le Bertre}, {Roussel}, \&
  {Turon}}]{epchetal85}
{Epchtein}, N., {Matsuura}, O.~T., {Braz}, M.~A., {et~al.} 1985, \aaps, 61, 203

\bibitem[{{Feast} {et~al.}(2006){Feast}, {Whitelock}, \&
  {Menzies}}]{feasetal06}
{Feast}, M.~W., {Whitelock}, P.~A., \& {Menzies}, J.~W. 2006, \mnras, 369, 791

\bibitem[{{Fong} {et~al.}(2006){Fong}, {Meixner}, {Sutton}, {Zalucha}, \&
  {Welch}}]{fongetal06}
{Fong}, D., {Meixner}, M., {Sutton}, E.~C., {Zalucha}, A., \& {Welch}, W.~J.
  2006, \apj, 652, 1626

\bibitem[{{Forestini} \& {Charbonnel}(1997)}]{forechar97}
{Forestini}, M. \& {Charbonnel}, C. 1997, \aaps, 123, 241

\bibitem[{{Gezari} {et~al.}(1987){Gezari}, {Schmitz}, \& {Mead}}]{gezaetal87}
{Gezari}, D.~Y., {Schmitz}, M., \& {Mead}, J.~M. 1987, {Catalog of infrared
  observations. Part 1: Data}, Tech. rep.

\bibitem[{{Gonz{\'a}lez Delgado} {et~al.}(2003){Gonz{\'a}lez Delgado},
  {Olofsson}, {Kerschbaum}, {Sch{\"o}ier}, {Lindqvist}, \&
  {Groenewegen}}]{delgetal03}
{Gonz{\'a}lez Delgado}, D., {Olofsson}, H., {Kerschbaum}, F., {et~al.} 2003,
  \aap, 411, 123

\bibitem[{{Groenewegen}(1997)}]{groe97}
{Groenewegen}, M.~A.~T. 1997, \aap, 317, 503

\bibitem[{{Groenewegen} {et~al.}(1993){Groenewegen}, {de Jong}, \&
  {Baas}}]{groeetal93}
{Groenewegen}, M.~A.~T., {de Jong}, T., \& {Baas}, F. 1993, \aaps, 101, 513

\bibitem[{{Groenewegen} {et~al.}(1992){Groenewegen}, {de Jong}, {van der
  Bliek}, {Slijkhuis}, \& {Willems}}]{groeetal92}
{Groenewegen}, M.~A.~T., {de Jong}, T., {van der Bliek}, N.~S., {Slijkhuis},
  S., \& {Willems}, F.~J. 1992, \aap, 253, 150

\bibitem[{{Groenewegen} {et~al.}(1998){Groenewegen}, {van der Veen}, \&
  {Matthews}}]{groeetal98}
{Groenewegen}, M.~A.~T., {van der Veen}, W.~E.~C.~J., \& {Matthews}, H.~E.
  1998, \aap, 338, 491

\bibitem[{{Habing}(1996)}]{habi96}
{Habing}, H.~J. 1996, \aapr, 7, 97

\bibitem[{{Habing} {et~al.}(1994){Habing}, {Tignon}, \& {Tielens}}]{habietal94}
{Habing}, H.~J., {Tignon}, J., \& {Tielens}, A.~G.~G.~M. 1994, \aap, 286, 523

\bibitem[{{Heras} \& {Hony}(2005)}]{herahony05}
{Heras}, A.~M. \& {Hony}, S. 2005, \aap, 439, 171

\bibitem[{{Heske} {et~al.}(1990){Heske}, {Forveille}, {Omont}, {van der Veen},
  \& {Habing}}]{hesketal90}
{Heske}, A., {Forveille}, T., {Omont}, A., {van der Veen}, W.~E.~C.~J., \&
  {Habing}, H.~J. 1990, \aap, 239, 173

\bibitem[{{H{\"o}fner} \& {Andersen}(2007)}]{hofnande07}
{H{\"o}fner}, S. \& {Andersen}, A.~C. 2007, \aap, 465, L39

\bibitem[{{H{\"o}fner} {et~al.}(1995){H{\"o}fner}, {Feuchtinger}, \&
  {Dorfi}}]{hofnetal95}
{H{\"o}fner}, S., {Feuchtinger}, M.~U., \& {Dorfi}, E.~A. 1995, \aap, 297, 815

\bibitem[{{Hogerheijde} \& {van der Tak}(2000)}]{hogevdta00}
{Hogerheijde}, M.~R. \& {van der Tak}, F.~F.~S. 2000, \aap, 362, 697

\bibitem[{{Ivezic} \& {Elitzur}(1997)}]{ivezelit97}
{Ivezic}, Z. \& {Elitzur}, M. 1997, \mnras, 287, 799

\bibitem[{{Jones} {et~al.}(1990){Jones}, {Bryja}, {Gehrz}, {Harrison},
  {Johnson}, {Klebe}, \& {Lawrence}}]{joneetal90}
{Jones}, T.~J., {Bryja}, C.~O., {Gehrz}, R.~D., {et~al.} 1990, \apjs, 74, 785

\bibitem[{{Justtanont} {et~al.}(1994){Justtanont}, {Skinner}, \&
  {Tielens}}]{justetal94}
{Justtanont}, K., {Skinner}, C.~J., \& {Tielens}, A.~G.~G.~M. 1994, \apj, 435,
  852

\bibitem[{{Justtanont} {et~al.}(1996){Justtanont}, {Skinner}, {Tielens},
  {Meixner}, \& {Baas}}]{justetal96}
{Justtanont}, K., {Skinner}, C.~J., {Tielens}, A.~G.~G.~M., {Meixner}, M., \&
  {Baas}, F. 1996, \apj, 456, 337

\bibitem[{{Justtanont} \& {Tielens}(1992)}]{justtiel92}
{Justtanont}, K. \& {Tielens}, A.~G.~G.~M. 1992, \apj, 389, 400

\bibitem[{{Kahane} {et~al.}(2000){Kahane}, {Dufour}, {Busso}, {Gallino},
  {Lugaro}, {Forestini}, \& {Straniero}}]{kahaetal00}
{Kahane}, C., {Dufour}, E., {Busso}, M., {et~al.} 2000, \aap, 357, 669

\bibitem[{{Kahane} \& {Jura}(1994)}]{kahajura94}
{Kahane}, C. \& {Jura}, M. 1994, \aap, 290, 183

\bibitem[{{Karakas} \& {Lattanzio}(2007)}]{karalatt07}
{Karakas}, A.~I. \& {Lattanzio}, J.~C. 2007, ArXiv e-prints, 708

\bibitem[{{Kastner}(1992)}]{kast92}
{Kastner}, J.~H. 1992, \apj, 401, 337

\bibitem[{{Kemper} {et~al.}(2003){Kemper}, {Stark}, {Justtanont}, {de Koter},
  {Tielens}, {Waters}, {Cami}, \& {Dijkstra}}]{kempetal03}
{Kemper}, F., {Stark}, R., {Justtanont}, K., {et~al.} 2003, \aap, 407, 609

\bibitem[{{Kerschbaum} \& {Hron}(1994)}]{kershron94}
{Kerschbaum}, F. \& {Hron}, J. 1994, \aaps, 106, 397

\bibitem[{{Knapp} \& {Morris}(1985)}]{knapmorr85}
{Knapp}, G.~R. \& {Morris}, M. 1985, \apj, 292, 640

\bibitem[{{Knapp} {et~al.}(1998){Knapp}, {Young}, {Lee}, \&
  {Jorissen}}]{knapetal98}
{Knapp}, G.~R., {Young}, K., {Lee}, E., \& {Jorissen}, A. 1998, \apjs, 117, 209

\bibitem[{{Lambert} {et~al.}(1986){Lambert}, {Gustafsson}, {Eriksson}, \&
  {Hinkle}}]{lambetal86}
{Lambert}, D.~L., {Gustafsson}, B., {Eriksson}, K., \& {Hinkle}, K.~H. 1986,
  \apjs, 62, 373

\bibitem[{{Lamers} \& {Cassinelli}(1999)}]{lamecass99}
{Lamers}, H.~J.~G.~L.~M. \& {Cassinelli}, J.~P. 1999, {Introduction to Stellar
  Winds} (Introduction to Stellar Winds, by Henny J.~G.~L.~M.~Lamers and Joseph
  P.~Cassinelli, pp.~452.~ISBN 0521593980.~Cambridge, UK: Cambridge University
  Press, June 1999.)

\bibitem[{{Le{\~a}o} {et~al.}(2006){Le{\~a}o}, {de Laverny}, {M{\'e}karnia},
  {de Medeiros}, \& {Vandame}}]{leaoetal06}
{Le{\~a}o}, I.~C., {de Laverny}, P., {M{\'e}karnia}, D., {de Medeiros}, J.~R.,
  \& {Vandame}, B. 2006, \aap, 455, 187

\bibitem[{{Le Bertre}(1992)}]{lebertre92}
{Le Bertre}, T. 1992, \aaps, 94, 377

\bibitem[{{Le Bertre}(1993)}]{lebertre93}
{Le Bertre}, T. 1993, \aaps, 97, 729

\bibitem[{{Lepine} {et~al.}(1995){Lepine}, {Ortiz}, \& {Epchtein}}]{lepietal95}
{Lepine}, J.~R.~D., {Ortiz}, R., \& {Epchtein}, N. 1995, \aap, 299, 453

\bibitem[{{Maercker} {et~al.}(2008){Maercker}, {Sch{\"o}ier}, {Olofsson},
  {Bergman}, \& {Ramstedt}}]{maeretal08}
{Maercker}, M., {Sch{\"o}ier}, F.~L., {Olofsson}, H., {Bergman}, P., \&
  {Ramstedt}, S. 2008, \aap, 479, 779

\bibitem[{{Mamon} {et~al.}(1988){Mamon}, {Glassgold}, \&
  {Huggins}}]{mamoetal88}
{Mamon}, G.~A., {Glassgold}, A.~E., \& {Huggins}, P.~J. 1988, \apj, 328, 797

\bibitem[{{Marengo} {et~al.}(1999){Marengo}, {Busso}, {Silvestro}, {Persi}, \&
  {Lagage}}]{mareetal99}
{Marengo}, M., {Busso}, M., {Silvestro}, G., {Persi}, P., \& {Lagage}, P.~O.
  1999, \aap, 348, 501

\bibitem[{{Marigo} \& {Girardi}(2007)}]{marigira07}
{Marigo}, P. \& {Girardi}, L. 2007, \aap, 469, 239

\bibitem[{{Marshall} {et~al.}(1992){Marshall}, {Leahy}, \& {Kwok}}]{marsetal92}
{Marshall}, C.~R., {Leahy}, D.~A., \& {Kwok}, S. 1992, \pasp, 104, 397

\bibitem[{{Mauron} \& {Huggins}(1999)}]{maurhugg99}
{Mauron}, N. \& {Huggins}, P.~J. 1999, \aap, 349, 203

\bibitem[{{Mauron} \& {Huggins}(2000)}]{maurhugg00}
{Mauron}, N. \& {Huggins}, P.~J. 2000, \aap, 359, 707

\bibitem[{{Mauron} \& {Huggins}(2001)}]{maurhugg01}
{Mauron}, N. \& {Huggins}, P.~J. 2001, in IAU Symposium, Vol. 205, Galaxies and
  their Constituents at the Highest Angular Resolutions, ed. R.~T. {Schilizzi},
  310--+

\bibitem[{{Mauron} \& {Huggins}(2006)}]{maurhugg06}
{Mauron}, N. \& {Huggins}, P.~J. 2006, \aap, 452, 257

\bibitem[{{Men'shchikov} {et~al.}(2006){Men'shchikov}, {Balega}, {Berger},
  {Driebe}, {Hofmann}, {Maximov}, {Schertl}, {Shenavrin}, \&
  {Weigelt}}]{mensetal06}
{Men'shchikov}, A.~B., {Balega}, Y.~Y., {Berger}, M., {et~al.} 2006, \aap, 448,
  271

\bibitem[{{Monnier} {et~al.}(1998){Monnier}, {Geballe}, \&
  {Danchi}}]{monnetal98}
{Monnier}, J.~D., {Geballe}, T.~R., \& {Danchi}, W.~C. 1998, \apj, 502, 833

\bibitem[{{Morris}(1975)}]{morris75}
{Morris}, M. 1975, \apj, 197, 603

\bibitem[{{Neri} {et~al.}(1998){Neri}, {Kahane}, {Lucas}, {Bujarrabal}, \&
  {Loup}}]{nerietal98}
{Neri}, R., {Kahane}, C., {Lucas}, R., {Bujarrabal}, V., \& {Loup}, C. 1998,
  \aaps, 130, 1

\bibitem[{{Olivier} {et~al.}(2001){Olivier}, {Whitelock}, \&
  {Marang}}]{olivetal01}
{Olivier}, E.~A., {Whitelock}, P., \& {Marang}, F. 2001, \mnras, 326, 490

\bibitem[{{Olofsson} {et~al.}(2000){Olofsson}, {Bergman}, {Lucas}, {Eriksson},
  {Gustafsson}, \& {Bieging}}]{olofetal00}
{Olofsson}, H., {Bergman}, P., {Lucas}, R., {et~al.} 2000, \aap, 353, 583

\bibitem[{{Olofsson} {et~al.}(1993){Olofsson}, {Eriksson}, {Gustafsson}, \&
  {Carlstrom}}]{olofetal93}
{Olofsson}, H., {Eriksson}, K., {Gustafsson}, B., \& {Carlstrom}, U. 1993,
  \apjs, 87, 267

\bibitem[{{Olofsson} {et~al.}(2002){Olofsson}, {Gonz{\'a}lez Delgado},
  {Kerschbaum}, \& {Sch{\"o}ier}}]{olofetal02}
{Olofsson}, H., {Gonz{\'a}lez Delgado}, D., {Kerschbaum}, F., \& {Sch{\"o}ier},
  F.~L. 2002, \aap, 391, 1053

\bibitem[{{Olofsson} {et~al.}(1982){Olofsson}, {Johansson}, {Hjalmarson}, \&
  {Nguyen-Quang-Rieu}}]{olofetal82}
{Olofsson}, H., {Johansson}, L.~E.~B., {Hjalmarson}, A., \&
  {Nguyen-Quang-Rieu}. 1982, \aap, 107, 128

\bibitem[{{Ramstedt} {et~al.}(2006){Ramstedt}, {Sch{\"o}ier}, {Olofsson}, \&
  {Lundgren}}]{ramsetal06}
{Ramstedt}, S., {Sch{\"o}ier}, F.~L., {Olofsson}, H., \& {Lundgren}, A.~A.
  2006, \aap, 454, L103

\bibitem[{{Sahai}(1990)}]{saha90}
{Sahai}, R. 1990, \apj, 362, 652

\bibitem[{{Samus} {et~al.}(2004){Samus}, {Durlevich}, \& {et al.}}]{samuetal04}
{Samus}, N.~N., {Durlevich}, O.~V., \& {et al.} 2004, VizieR Online Data
  Catalog, 2250, 0

\bibitem[{{Sandin} \& {H{\"o}fner}(2003)}]{sandhofn03}
{Sandin}, C. \& {H{\"o}fner}, S. 2003, \aap, 398, 253

\bibitem[{{Sch{\"o}ier} {et~al.}(2005{\natexlab{a}}){Sch{\"o}ier}, {Lindqvist},
  \& {Olofsson}}]{schoetal05b}
{Sch{\"o}ier}, F.~L., {Lindqvist}, M., \& {Olofsson}, H. 2005{\natexlab{a}},
  \aap, 436, 633

\bibitem[{{Sch{\"o}ier} \& {Olofsson}(2001)}]{schoolof01}
{Sch{\"o}ier}, F.~L. \& {Olofsson}, H. 2001, \aap, 368, 969

\bibitem[{{Sch{\"o}ier} {et~al.}(2002){Sch{\"o}ier}, {Ryde}, \&
  {Olofsson}}]{schoetal02}
{Sch{\"o}ier}, F.~L., {Ryde}, N., \& {Olofsson}, H. 2002, \aap, 391, 577

\bibitem[{{Sch{\"o}ier} {et~al.}(2005{\natexlab{b}}){Sch{\"o}ier}, {van der
  Tak}, {van Dishoeck}, \& {Black}}]{schoetal05}
{Sch{\"o}ier}, F.~L., {van der Tak}, F.~F.~S., {van Dishoeck}, E.~F., \&
  {Black}, J.~H. 2005{\natexlab{b}}, \aap, 432, 369

\bibitem[{{Schr{\"o}der} \& {Sedlmayr}(2001)}]{schrsedl01}
{Schr{\"o}der}, K.-P. \& {Sedlmayr}, E. 2001, \aap, 366, 913

\bibitem[{{Sedlmayr}(1994)}]{sedl94}
{Sedlmayr}, E. 1994, in Lecture Notes in Physics, Berlin Springer Verlag, Vol.
  428, IAU Colloq. 146: Molecules in the Stellar Environment, ed. U.~G.
  {Jorgensen}, 163--+

\bibitem[{{Simis} {et~al.}(2001){Simis}, {Icke}, \& {Dominik}}]{simietal01}
{Simis}, Y.~J.~W., {Icke}, V., \& {Dominik}, C. 2001, \aap, 371, 205

\bibitem[{{Skinner} {et~al.}(1997){Skinner}, {Meixner}, {Barlow}, {Collison},
  {Justtanont}, {Blanco}, {Pina}, {Ball}, {Keto}, {Arens}, \&
  {Jernigan}}]{skinetal97}
{Skinner}, C.~J., {Meixner}, M., {Barlow}, M.~J., {et~al.} 1997, \aap, 328, 290

\bibitem[{{Sopka} {et~al.}(1985){Sopka}, {Hildebrand}, {Jaffe}, {Gatley},
  {Roellig}, {Werner}, {Jura}, \& {Zuckerman}}]{sopketal85}
{Sopka}, R.~J., {Hildebrand}, R., {Jaffe}, D.~T., {et~al.} 1985, \apj, 294, 242

\bibitem[{{Speck} {et~al.}(1997){Speck}, {Barlow}, \& {Skinner}}]{specetal97}
{Speck}, A.~K., {Barlow}, M.~J., \& {Skinner}, C.~J. 1997, \mnras, 288, 431

\bibitem[{{Stanek} {et~al.}(1995){Stanek}, {Knapp}, {Young}, \&
  {Phillips}}]{stanetal95}
{Stanek}, K.~Z., {Knapp}, G.~R., {Young}, K., \& {Phillips}, T.~G. 1995, \apjs,
  100, 169

\bibitem[{{Steffen} \& {Sch{\"o}nberner}(2000)}]{stefscho00}
{Steffen}, M. \& {Sch{\"o}nberner}, D. 2000, \aap, 357, 180

\bibitem[{{Suh}(2000)}]{suh00}
{Suh}, K.-W. 2000, \mnras, 315, 740

\bibitem[{{Teyssier} {et~al.}(2006){Teyssier}, {Hernandez}, {Bujarrabal},
  {Yoshida}, \& {Phillips}}]{teysetal06}
{Teyssier}, D., {Hernandez}, R., {Bujarrabal}, V., {Yoshida}, H., \&
  {Phillips}, T.~G. 2006, \aap, 450, 167

\bibitem[{{van Loon}(2006)}]{loon06}
{van Loon}, J.~T. 2006, ArXiv Astrophysics e-prints

\bibitem[{{Volk} \& {Cohen}(1989)}]{volkcohe89}
{Volk}, K. \& {Cohen}, M. 1989, \aj, 98, 1918

\bibitem[{{Walmsley} {et~al.}(1991){Walmsley}, {Chini}, {Kreysa}, {Steppe},
  {Forveille}, \& {Omont}}]{walmetal91}
{Walmsley}, C.~M., {Chini}, R., {Kreysa}, E., {et~al.} 1991, \aap, 248, 555

\bibitem[{{Whitelock} {et~al.}(1994){Whitelock}, {Menzies}, {Feast}, {Marang},
  {Carter}, {Roberts}, {Catchpole}, \& {Chapman}}]{whitetal94}
{Whitelock}, P., {Menzies}, J., {Feast}, M., {et~al.} 1994, \mnras, 267, 711

\bibitem[{{Whittet}(2003)}]{whit03}
{Whittet}, D.~C.~B., ed. 2003, {Dust in the galactic environment}

\bibitem[{{Willson}(2000)}]{will00}
{Willson}, L.~A. 2000, \araa, 38, 573

\bibitem[{{Winters} {et~al.}(2003){Winters}, {Le Bertre}, {Jeong}, {Nyman}, \&
  {Epchtein}}]{wintetal03}
{Winters}, J.~M., {Le Bertre}, T., {Jeong}, K.~S., {Nyman}, L.-{\AA}., \&
  {Epchtein}, N. 2003, \aap, 409, 715

\bibitem[{{Woitke}(2006)}]{woit06}
{Woitke}, P. 2006, \aap, 460, L9

\bibitem[{{Woods} {et~al.}(2003){Woods}, {Sch{\"o}ier}, {Nyman}, \&
  {Olofsson}}]{woodetal03}
{Woods}, P.~M., {Sch{\"o}ier}, F.~L., {Nyman}, L.-{\AA}., \& {Olofsson}, H.
  2003, \aap, 402, 617

\bibitem[{{Zuckerman} \& {Dyck}(1986)}]{zuckdyck86}
{Zuckerman}, B. \& {Dyck}, H.~M. 1986, \apj, 304, 394

\end{thebibliography}

\clearpage

\appendix

\section{A new mass loss rate formula}

\citet{knapmorr85} derived a theoretical mass-loss-rate formula in terms of the CO($J$\,=\,1\,$\rightarrow$\,0) line intensity, valid for optically thick CO emission. The formula relates the mass-loss rate to a number of easily determined observables; $T_{mb}$ for the CO(1-0) line, and $v_{\rm{e}}$. The distance D and CO abundance $f_{\rm{CO}}$ also enter the formula. It was derived using a fixed CO envelope size.This formula was used in e. g. \citet{olofetal93} to estimate mass-loss rates for a sample of carbon stars. Later, \citet{schoolof01} could compare their results on the sample, derived using the same radiative transfer model as in this work, to those estimated using the formula. They found that the formula underestimated the mass-loss rates compared to what was found from the radiative transfer analysis, on average by a factor of four, but the discrepancy was found to increase even more for lower mass-loss rate objects. The reason is that the emission from the low-mass-loss-rate objects is not optically thick and it is more sensitive to the size of the CO envelope. However, the usefulness of simple formulae of this sort can not be disputed.

In order to improve the results by \citet{knapmorr85}, we set up a grid with 60 model stars, whose properties were varied in $\dot{M}$ (1, 3, 10, 30, 
and 100\,$\times$\,10$^{-7}$\,M$_\odot$\,yr$^{-1}$), $v_{\mathrm{e}}$ (5, 10, 15, and 20\,km\,s$^{-1}$), 
and $f_{\mathrm{CO}}$ (1, 3, and 10\,$\times$\,10$^{-4}$). Following \citet{schoolof01}, the value of $h$ was assumed to be 0.2 for $\dot{M}$ in the range up to 3\,$\times$\,10$^{-7}$\,M$_\odot$\,yr$^{-1}$, 0.5 for 1 -- 3\,$\times$\,10$^{-6}$\,M$_\odot$\,yr$^{-1}$, and 1.5 for the mass-loss rate 1\,$\times$\,10$^{-5}$\,M$_\odot$\,yr$^{-1}$. For each star, the velocity integrated intensities in CO ($J$\,=\,1\,$\rightarrow$\,0, $J$\,=\,2\,$\rightarrow$\,1, $J$\,=\,3\,$\rightarrow$\,2, $J$\,=\,4\,$\rightarrow$\,3) were calculated for a 20\,m class telescope (the corresponding beam sizes were 33$\arcsec$, 16.5$\arcsec$, 11$\arcsec$, and 8.5$\arcsec$). The distance to the stars were set to 1\,kpc. The output from the modeling was fitted to Eqn.~(\ref{MdotEq}) and the errors were 
minimized using a Levenberg-Marquardt method (Press et al. 1992),

\begin{equation}
\label{MdotEq}
\dot{M}=s_{J}(I_{\mathrm{CO}}\theta^{2}_\mathrm{b}D^{2})^{a_J}v_{\mathrm{e}}^{b_J}f_{\mathrm{CO}}^{-c_J}.
\end{equation}

\begin{figure}
\centering{   
\includegraphics[width=9cm]{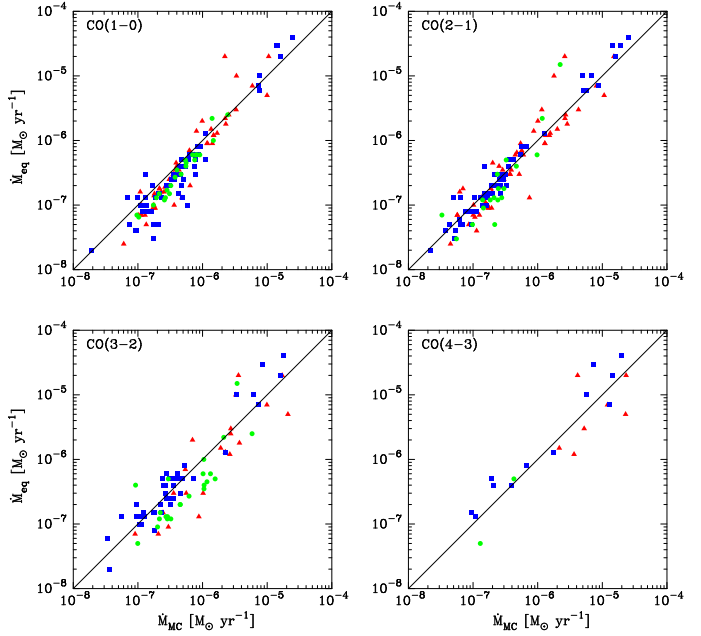}
\caption{Mass-loss rates derived using Eqn.~\ref{MdotEq} and the parameters in Table~\ref{MdotPara} versus the mass-loss rates estimated using the Monte Carlo model for the same stars. Red triangles represent carbon stars, blue squares, M-type stars, and green dots represent S-type stars.}
\label{turb}}
\end{figure}

Table~\ref{MdotPara} 
summarizes the results. For low upper-$J$ levels, the mass-loss rate is about equally (and moderately)
dependent on changes of $I$, $v_{\mathrm{e}}$, and $f_{\mathrm{CO}}$. At high upper-$J$ levels, the mass- 
loss rate is almost linearly dependent on the velocity-integrated intensity ($I$), but 
relatively insensitive to changes in $v_{\mathrm{e}}$, and $f_{\mathrm{CO}}$. In general, the parameter 
most difficult to fit was the exponent for the expansion velocity ($v_{\mathrm{e}}$), while the 
exponent for the velocity-integrated intensity ($I$) is well determined. Eqn.~(\ref{MdotEq}) only applies to unresolved emission. Fig.~\ref{turb} shows a comparison between mass-loss rates estimated using Eqn.~(\ref{MdotEq}) on the y-axis and mass-loss rates estimated using the Monte-Carlo radiative transfer model on the x-axis for three samples of C-, M-, and S-stars \citep{schoolof01,olofetal02,delgetal03,ramsetal06}. $f_{\rm{CO}}$ is assumed to be $1\times 10^{-3}$ for the C-stars, $2\times 10^{-4}$ for the M-stars, and $6\times 10^{-4}$ for the S-stars.

\begin{table}[h]
\caption{Values for the parameters in Eq.~\ref{MdotEq}. $s_J$ is calibrated such that Eq.~\ref{MdotEq} gives the mass-loss rate in M$_\odot$\,yr$^{-1}$. Values are valid in the ranges 10$^{-7}$\,$\leq$\,$\dot{M}$\,$\leq$\,10$^{-5}$\,M$_\odot$\,yr$^{-1}$, 5\,$\leq$\,$v_{\mathrm{e}}$\,$\leq$\,20\,km\,s$^{-1}$, 10$^{-4}$\,$\leq$\,$f_{\mathrm{CO}}$\,$\leq$\,10$^{-3}$.}
\label{MdotPara}
$
\begin{array}{p{0.15\linewidth}cccccc}
\hline
\noalign{\smallskip}
\multicolumn{1}{l}{{\mathrm{transition}}} &
\multicolumn{1}{c}{s_J} &
\multicolumn{1}{c}{a_J\ (I)} & 
\multicolumn{1}{c}{b_J\ (v_{\mathrm{e}})} &
\multicolumn{1}{c}{c_J\ (f_{\mathrm{CO}})} \\ 
\noalign{\smallskip}
\hline
\noalign{\smallskip}
$J$\,=\,1\,$\rightarrow$\,0  & 4.9(\pm2.4)\times10^{-12} & 0.68(\pm0.03) & 0.59(\pm0.13) & 0.80(\pm0.08) \\
$J$\,=\,2\,$\rightarrow$\,1  & 1.3(\pm0.7)\times10^{-11} & 0.82(\pm0.03) & 0.46(\pm0.13) & 0.59(\pm0.07) \\
$J$\,=\,3\,$\rightarrow$\,2  & 3.8(\pm3.4)\times10^{-11} & 0.91(\pm0.04) & 0.39(\pm0.13) & 0.45(\pm0.07) \\
$J$\,=\,4\,$\rightarrow$\,3  & 8.4(\pm4.1)\times10^{-11} & 0.95(\pm0.04) & 0.36(\pm0.13) & 0.37(\pm0.07) \\
\noalign{\smallskip}
\hline
\end{array}
$
\end{table}

\clearpage

\onecolumn

\section{Photometric fluxes and dust radiative transfer}
\label{a:flux}

The SEDs used as constraints in the dust emission modelling are constructed from JHKLM-band, sub-millimetre, and millimetre photometric flux densities from the literature, together with IRAS data. All flux densities used are listed in Table~\ref{fluxes}. Values in italics were omitted from the analysis to avoid a bias in the $\chi^{2}$-analysis at certain wavelengths. Values in italics marked with a dagger were omitted since the reported errors were $\gtrsim$\,50\%.

\begin{table*}[h]
\caption{Photometric flux densities in 10$^{-26}$ W\,m$^{-2}$\,Hz$^{-1}$.}
\label{fluxes}
$$
\begin{array}{c|cccccccccccccccccccc}
\hline
\noalign{\smallskip}
\multicolumn{1}{c|}{{\mathrm{wavelength}}} &&
\multicolumn{1}{c}{{\mathrm{LP\ And}}} &&
\multicolumn{1}{c}{{\mathrm{CW\ Leo}}} &&
\multicolumn{1}{c}{{\mathrm{V384\ Per}}} &&
\multicolumn{1}{c}{{\mathrm{AFGL\ 3068}}} &&
\multicolumn{1}{c}{{\mathrm{RW\ LMi}}} &&
\multicolumn{1}{c}{{\mathrm{TX\ Cam}}} &&
\multicolumn{1}{c}{{\mathrm{GX\ Mon}}} &&
\multicolumn{1}{c}{{\mathrm{WX\ Psc}}} &&
\multicolumn{1}{c}{{\mathrm{IK\ Tau}}} &&
\multicolumn{1}{c}{{\mathrm{IRC}-10529}} \\
\multicolumn{1}{c|}{[\mathrm{\mu m}]} && && && && && && && && && \\
\hline
\noalign{\smallskip}
1.24 && 1.6^{\rm{a}} && 4.0^{\rm{d}} && 13^{\rm{e}} && \cdots && 5.8^{\rm{h}} && 280^{\rm{a}} && 49^{\rm{j}} && 3.2^{\rm{i}} && 250^{\rm{i}} && 2.4^{\rm{i}} \\ 
1.63 && 11^{\rm{a}} && 58^{\rm{d}} && 53^{\rm{e}} && \cdots && 39^{\rm{h}} && 590^{\rm{a}} && 120^{\rm{j}} && 32^{\rm{i}} && 730^{\rm{i}} && 20^{\rm{i}} \\ 
2.19 && 55^{\rm{a}} && 520^{\rm{d}} && 170^{\rm{e}} && \cdots && 180^{\rm{h}} && 780^{\rm{a}} && 200^{\rm{j}} && 130^{\rm{i}} && 1200^{\rm{i}} && 78^{\rm{i}} \\ 
3.79 && 450^{\rm{a}} && 10000^{\rm{d}} && 520^{\rm{e}} && 9.6^{\rm{g}} && 970^{\rm{h}} && 750^{\rm{a}} && 580^{\rm{j}} && 620^{\rm{i}} && 1900^{\rm{i}} && 290^{\rm{i}} \\ 
4.64 && \cdots && 19000^{\rm{d}} && \cdots && 55^{\rm{g}} && \cdots && \cdots && 410^{\rm{j}} && 800^{\rm{i}} && 1800^{\rm{i}} && 340^{\rm{i}} \\ 
8.55 && \cdots && \cdots && \cdots && 290^{\rm{k}} && \cdots && \cdots && \cdots && {\it 850}^{\rm{k}} && \cdots && \cdots \\ 
8.57 && \cdots && \cdots && \cdots && \cdots && \cdots && \cdots && \cdots && {\it 700}^{\rm{k}} && \cdots && \cdots \\ 
9.56 && \cdots && \cdots && \cdots &&  {\it 360}^{\rm{k}} && \cdots && \cdots && \cdots &&  {\it 1100}^{\rm{k}} && \cdots && \cdots \\ 
11.21 && \cdots && \cdots && \cdots &&  {\it 400}^{\rm{k}} &&  {\it 3900}^{\rm{k}} && \cdots && \cdots &&  {\it 1000}^{\rm{k}} && \cdots && \cdots \\ 
12 && 960^{\rm{b}} && 48000^{\rm{b}} && 540^{\rm{b}} && 1700^{\rm{b}} && 3300^{\rm{b}} && 1600^{\rm{b}} && 1600^{\rm{b}} && 1200^{\rm{b}} && 4600^{\rm{b}} && 1300^{\rm{b}} \\ 
12.45 && \cdots && \cdots && \cdots &&  {\it 470}^{\rm{k}} && \cdots && \cdots && \cdots && {\it 360}^{\rm{k}} && \cdots && \cdots \\ 
25 && 470^{\rm{b}} && 23000^{\rm{b}} && 200^{\rm{b}} && 780^{\rm{b}} && 1200^{\rm{b}} && 640^{\rm{b}} && 360^{\rm{b}} && 970^{\rm{b}} && 2400^{\rm{b}} && 1100^{\rm{b}} \\ 
60 && 110^{\rm{b}} && 5700^{\rm{b}} && 40^{\rm{b}} && 250^{\rm{b}} && 270^{\rm{b}} && 130^{\rm{b}} && 110^{\rm{b}} && 220^{\rm{b}} && 330^{\rm{b}} && 220^{\rm{b}} \\
100 &&  36^{\rm{b}} && 920^{\rm{b}} && 12^{\rm{b}} && 74^{\rm{b}} && 86^{\rm{b}} && 39^{\rm{b}} && 40^{\rm{b}} && 72^{\rm{b}} && 100^{\rm{b}} && 64^{\rm{b}} \\  
400 && \cdots && \cdots && \cdots && \cdots && \cdots && \cdots && \cdots && 3.0^{\rm{m}} && \cdots && \cdots \\ 
450 && {\it 0.35^{\rm{f} \dagger}} && 13^{\rm{o}} && \cdots && 2.2^{\rm{o}} && 2.4^{\rm{o}} && \cdots && \cdots && \cdots && {\it 1.2}^{\rm{l} \dagger} && \cdots \\  
600 && 0.41^{\rm{f}} && \cdots &&  {\it 0.11}^{\rm{f} \dagger} && 0.67^{\rm{f}} && \cdots && \cdots && \cdots && \cdots && \cdots && \cdots \\ 
800 && 0.25^{\rm{f}} && {\it 6.7}^{\rm{l}} && 0.21^{\rm{f}} && {\it 0.33}^{\rm{f}} && 0.78^{\rm{l}} && \cdots && \cdots && \cdots && 0.25^{\rm{l}} && \cdots \\ 
850 && \cdots && 12^{\rm{o}} && \cdots && 0.59^{\rm{o}} && {\it 0.97}^{\rm{o}} && \cdots && \cdots && \cdots && \cdots && \cdots \\ 
1100 && 0.081^{\rm{f}} && {\it 1.9}^{\rm{l}} && 0.045^{\rm{f}} && 0.17^{\rm{f}} && 0.16^{\rm{l}} && \cdots && \cdots && \cdots && 0.082^{\rm{l}} && \cdots \\ 
1200 && \cdots && \cdots && \cdots && \cdots && \cdots && \cdots && \cdots && 0.24^{\rm{c}} && \cdots && 0.10^{\rm{c}} \\ 
1300 && \cdots && 1.5^{\rm{n}} && \cdots && \cdots && {\it 0.29}^{\rm{n}} && \cdots && \cdots && 0.059^{\rm{c}} && 0.037^{\rm{n}} && \cdots \\ 
\noalign{\smallskip}
\hline
\end{array}
$$
\medskip
{\bf References:} \\
$^{\rm{a}}$ \cite{bagnulo96} \\
$^{\rm{b}}$ \cite{iras88}, IRAS Point Source Catalogue \\
$^{\rm{c}}$ \cite{dehaetal07} \\
$^{\rm{d}}$ \cite{epchetal85} \\
$^{\rm{e}}$ \cite{gezaetal87} \\
$^{\rm{f}}$ \cite{groeetal93} \\
$^{\rm{g}}$ \cite{joneetal90} \\
$^{\rm{h}}$ \cite{kershron94} \\
$^{\rm{i}}$ \cite{lebertre93} \\
$^{\rm{j}}$ \cite{lepietal95} \\
$^{\rm{k}}$ \cite{mareetal99} \\
$^{\rm{l}}$ \cite{marsetal92} \\
$^{\rm{m}}$ \cite{sopketal85} \\
$^{\rm{n}}$ \cite{walmetal91} \\
$^{\rm{o}}$ JCMT archival data 
\end{table*}

\newpage

\section{Results from the CO line radiative transfer model}
\label{a:CO}

\begin{figure*}[h]
\raggedright 
{\includegraphics[width=15.0cm]{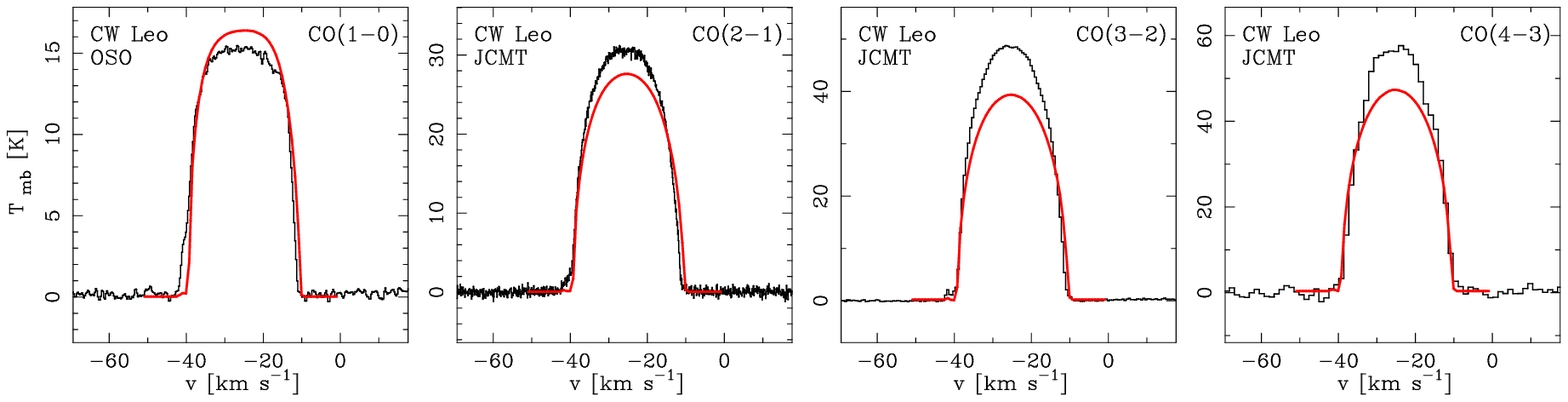}}
\caption{The observed CO $J$\,=\,1\,$\rightarrow$\,0 line from OSO, 2\,$\rightarrow$\,1, 3\,$\rightarrow$\,2, and 4\,$\rightarrow$\,3 lines from JCMT, of the carbon star CW Leo, overlayed with the best-fit model ($\dot{M}$=2$\,\times\,$10$^{-5}$\,M$_{\odot}$\,yr$^{-1}$) from the CO radiative transfer modelling. The reduced $\chi^{2}$ for this model is $\chi^{2}_{\rm{red}}$=0.6. See Sect.~\ref{sss:cwleo}.}
\label{cwleo_mod}
\vspace{0.2cm}
{\includegraphics[width=15.0cm]{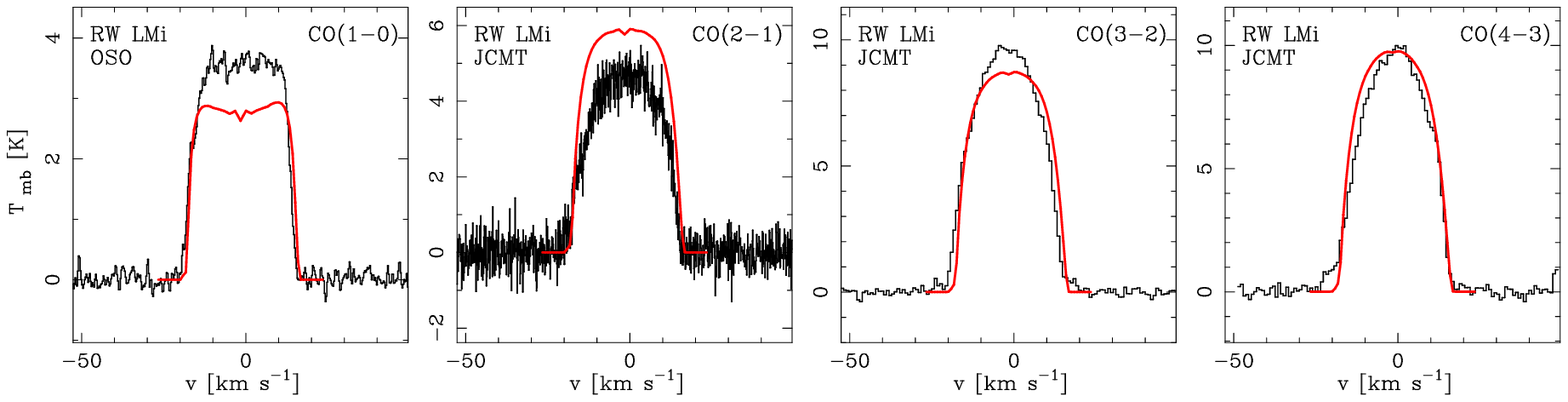}}
\caption{The observed CO $J$\,=\,1\,$\rightarrow$\,0 line from OSO, 2\,$\rightarrow$\,1, 3\,$\rightarrow$\,2, and 4\,$\rightarrow$\,3 lines from JCMT, of the carbon star RW LMi, overlayed with the best-fit model ($\dot{M}$=5$\,\times\,$10$^{-6}$\,M$_{\odot}$\,yr$^{-1}$) from the CO radiative transfer modelling. The reduced $\chi^{2}$ for this model is $\chi^{2}_{\rm{red}}$=2.1. See Sect.~\ref{sss:rwlmi}.}
\label{rwlmi_mod}
\vspace{0.2cm}
{\includegraphics[width=15.0cm]{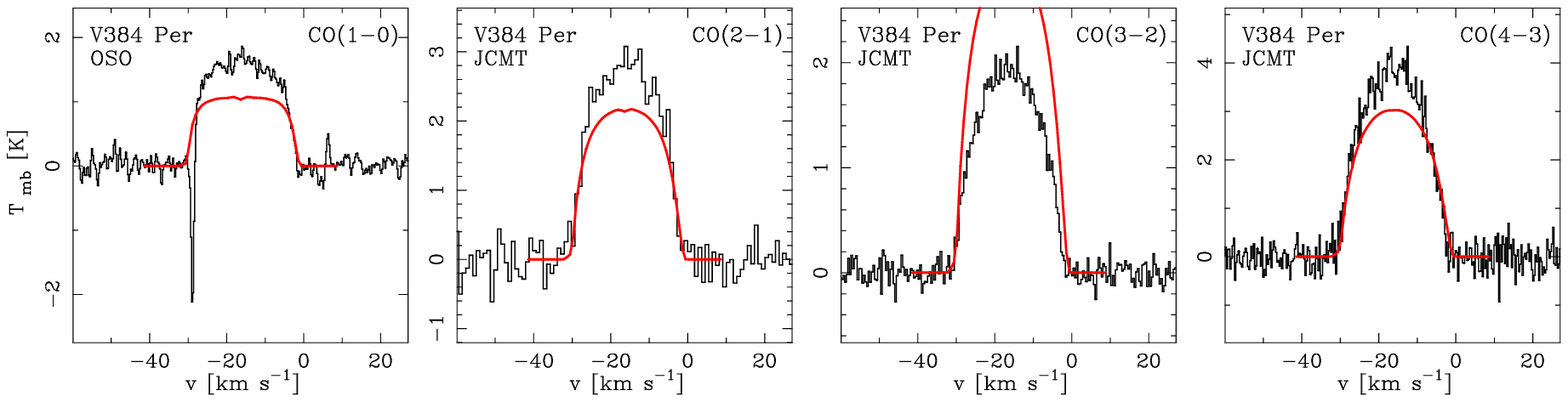}}
\caption{The observed CO $J$\,=\,1\,$\rightarrow$\,0 line from OSO, 2\,$\rightarrow$\,1, 3\,$\rightarrow$\,2, and 4\,$\rightarrow$\,3 lines from JCMT, of the carbon star V384 Per, overlayed with the best-fit model ($\dot{M}$=3$\,\times\,$10$^{-6}$\,M$_{\odot}$\,yr$^{-1}$) from the CO radiative transfer modelling. The reduced $\chi^{2}$ for this model is $\chi^{2}_{\rm{red}}$=3.7. See Sect.~\ref{sss:v384per}.}
\label{v384per_mod}
\vspace{0.2cm}
{\includegraphics[width=18.5cm]{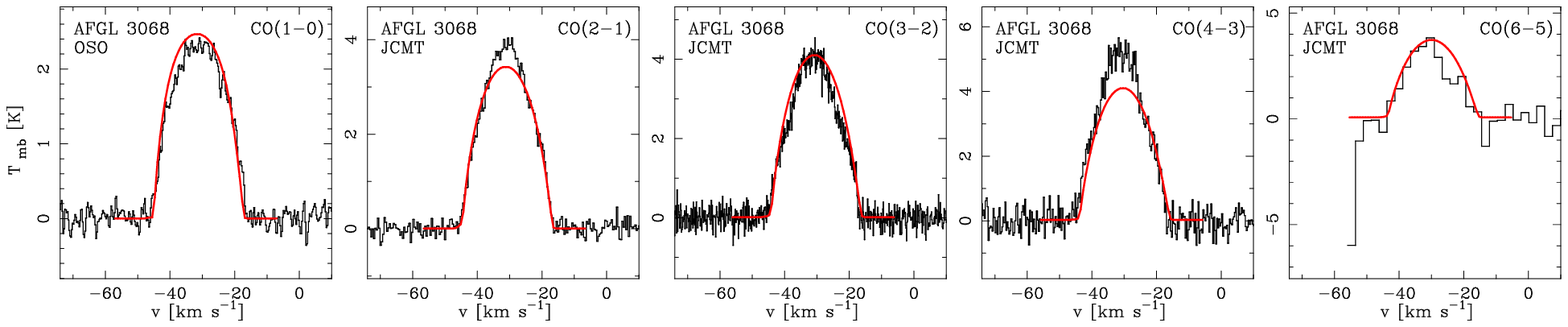}}
\caption{The observed CO $J$\,=\,1\,$\rightarrow$\,0 line from OSO, 2\,$\rightarrow$\,1, 3\,$\rightarrow$\,2, 4\,$\rightarrow$\,3, and 6\,$\rightarrow$\,5 lines from JCMT, of the carbon star AFGL 3068, overlayed with the best-fit model ($\dot{M}$=2$\,\times\,$10$^{-5}$\,M$_{\odot}$\,yr$^{-1}$) from the CO radiative transfer modelling. The reduced $\chi^{2}$ for this model is $\chi^{2}_{\rm{red}}$=0.7. See Sect.~\ref{sss:afgl3068}.}
\label{afgl3068_mod}
\end{figure*}

\begin{figure*}[t]
\raggedright 
\vspace{0.2cm}
{\includegraphics[width=15.0cm]{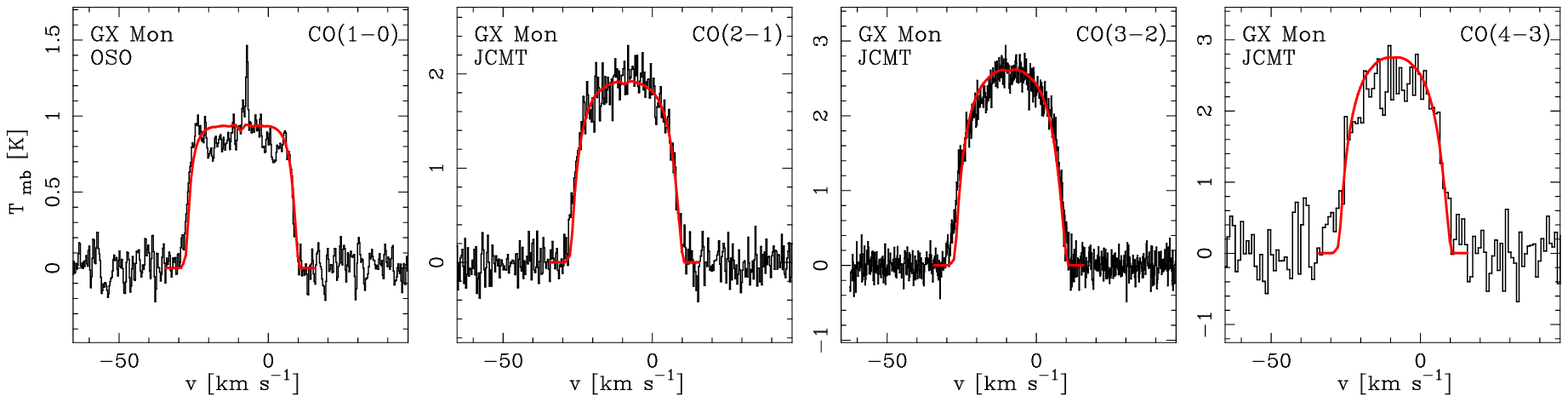}}
\caption{The observed CO $J$\,=\,1\,$\rightarrow$\,0 line from OSO, 2\,$\rightarrow$\,1, 3\,$\rightarrow$\,2, and 4\,$\rightarrow$\,3 lines from JCMT, of the M-type star GX Mon, overlayed with the best-fit model ($\dot{M}$=2$\,\times\,$10$^{-5}$\,M$_{\odot}$\,yr$^{-1}$) from the CO radiative transfer modelling. The reduced $\chi^{2}$ for this model is $\chi^{2}_{\rm{red}}$=0.04. See Sect.~\ref{sss:gxmon}.}
\label{gxmon_mod}
\vspace{0.2cm}
{\includegraphics[width=18.5cm]{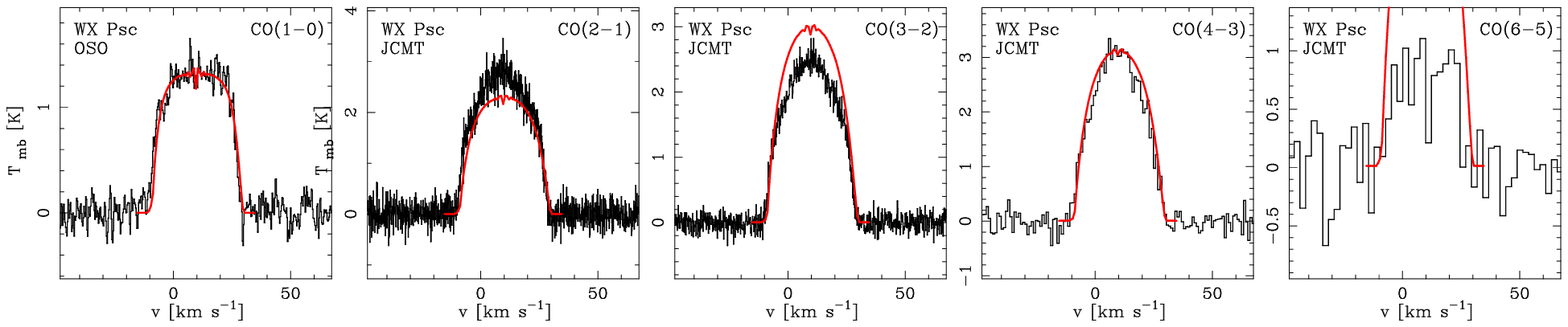}}
\caption{The observed CO $J$\,=\,1\,$\rightarrow$\,0 line from OSO, 2\,$\rightarrow$\,1, 3\,$\rightarrow$\,2, 4\,$\rightarrow$\,3, and 6\,$\rightarrow$\,5 lines from JCMT, of the M-type star WX Psc, overlayed with the best-fit model ($\dot{M}$=4$\,\times\,$10$^{-5}$\,M$_{\odot}$\,yr$^{-1}$) from the CO radiative transfer modelling. The reduced $\chi^{2}$ for this model is $\chi^{2}_{\rm{red}}$=1.0, when the 6\,$\rightarrow$\,5 line is omitted. See Sect.~\ref{sss:wxpsc}.}
\label{wxpsc_mod}
\vspace{0.2cm}
{\includegraphics[width=15.0cm]{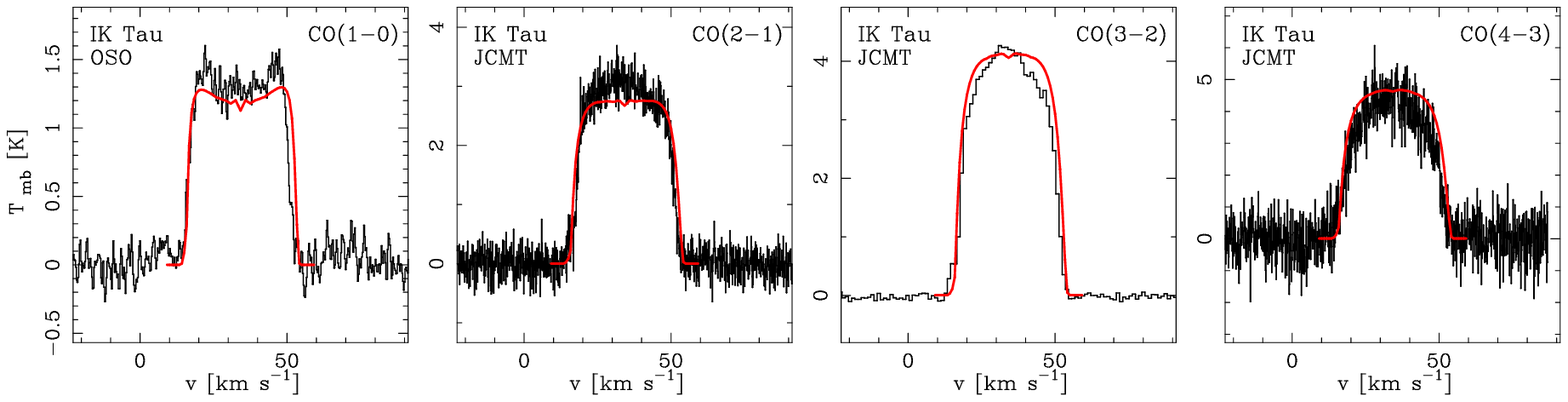}}
\caption{The observed CO $J$\,=\,1\,$\rightarrow$\,0 line from OSO, 2\,$\rightarrow$\,1, 3\,$\rightarrow$\,2, and 4\,$\rightarrow$\,3 lines from JCMT, of the {\bf M-type} star IK Tau, overlayed with the best-fit model ($\dot{M}$=1$\,\times\,$10$^{-5}$\,M$_{\odot}$\,yr$^{-1}$) from the CO radiative transfer modelling. The reduced $\chi^{2}$ for this model is $\chi^{2}_{\rm{red}}$=0.5. See Sect.~\ref{sss:iktau}.}
\label{iktau_mod}
\vspace{0.2cm}
{\includegraphics[width=15.0cm]{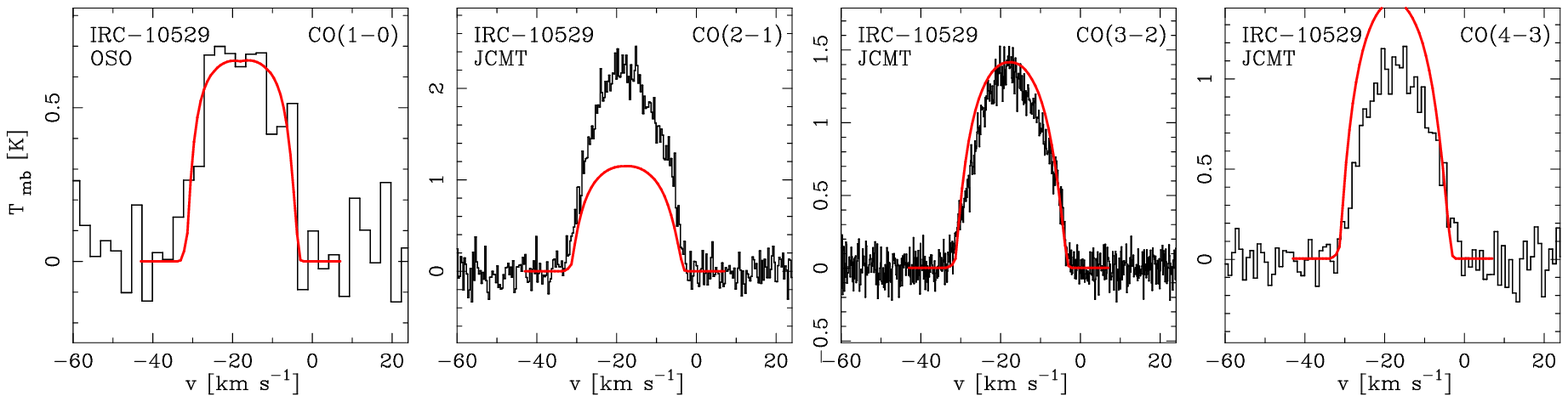}}
\caption{The observed CO $J$\,=\,1\,$\rightarrow$\,0 line from OSO, 2\,$\rightarrow$\,1, 3\,$\rightarrow$\,2, and 4\,$\rightarrow$\,3 lines from JCMT, of the {\bf M-type}  star IRC-10529, overlayed with the best-fit model ($\dot{M}$=1.6$\,\times\,$10$^{-5}$\,M$_{\odot}$\,yr$^{-1}$) from the CO radiative transfer modelling. The reduced $\chi^{2}$ for this model is $\chi^{2}_{\rm{red}}$=4.6. See Sect.~\ref{sss:irc10529}.}
\label{irc10529_mod}
\end{figure*}

Figs ~\ref{cwleo_mod}-\ref{irc10529_mod} show the best-fit models from the CO radiative transfer modelling overlayed the new observational data for all stars in the sample (except LP And and TX Cam shown in Figs~\ref{lpand_mod} and \ref{txcam_mod}, respectively). The results are discussed in Sect.~\ref{s:ind}. Table~\ref{intensities_diff} gives the difference between in integrated intensity from the model and the observation, for each line.

\begin{table}[h]
\caption{The difference between the integrated CO($J\rightarrow J-1$) intensities from the best-fit model and the observations, in percent; $(I_{\rm{mod}}-I_{\rm{obs}})/I_{\rm{obs}}$. }
\label{intensities_diff}
$$
\begin{array}{p{0.27\linewidth}cccccccc}
\hline
\noalign{\smallskip}
\multicolumn{1}{l}{{\mathrm{Source}}} &
\multicolumn{1}{c}{{\mathrm{OSO}}} &&
\multicolumn{4}{c}{{\mathrm{JCMT}}} \\ 
\cline{2-2}
\cline{4-7}
&
\multicolumn{1}{c}{1\rightarrow0} &&
\multicolumn{1}{c}{2\rightarrow1} &
\multicolumn{1}{c}{3\rightarrow2} &
\multicolumn{1}{c}{4\rightarrow3}& 
\multicolumn{1}{c}{6\rightarrow5} \\
\noalign{\smallskip}
\hline
\noalign{\smallskip}
\object{LP And} & \phantom{00}3.2 && \phantom{00}5.4 & \phantom{0}15.8 &  \phantom{00}4.2 & \phantom{0}$-$0.6 \\
\object{CW Leo}	 & \phantom{00}4.8 && \phantom{0}$-$7.3  & $-$15.6  & $-$11.7 & \cdots \\
\object{RW LMi} & $-$16.6 && \phantom{0}37.4 &  \phantom{0}$-$1.6 &  \phantom{00}6.2 & \cdots \\
\object{V384 Per} & $-$22.3 &&  $-$20.3 & \phantom{0}63.8  & $-$15.3 & \cdots \\
\object{AFGL 3068} & \phantom{00}6.3 &&  \phantom{0}$-$4.2  &  \phantom{0}11.0 & $-$13.3 &  \phantom{0}34.6 \\	
 & && & & & \\
\object{TX Cam} & \phantom{0}$-$7.7 && \phantom{0}56.9  & \phantom{0}40.2 & $-$18.9 & \cdots \\
\object{GX Mon}& \phantom{00}1.6 &&  \phantom{0}$-$3.4 &  \phantom{0}$-$1.9 & \phantom{00}3.4 & \cdots \\
\object{WX Psc} & \phantom{0}$-$3.2 && $-$14.8 &  \phantom{0}23.8 & \phantom{00}6.8  & 273.7  \\	
\object{IK Tau}	& \phantom{0}$-$3.3 && \phantom{0}$-$1.4 & \phantom{00}9.2 &\phantom{0}16.4 & \cdots \\
\object{IRC--10529}	& \phantom{0}$-$3.5 && $-$44.8 & \phantom{0}11.9  & \phantom{0}38.9 & \cdots \\
\noalign{\smallskip}
\hline
\end{array}
$$
\end{table}

\end{document}